\documentclass[structabstract]{aa} 
\usepackage{amstext,amsmath,mathtools}
\usepackage{graphicx}
\usepackage{textcomp}
\usepackage{txfonts}
\usepackage{time}
\usepackage{subfig}
\usepackage{color}
\usepackage{url}
\usepackage{natbib}
\bibpunct{(}{)}{;}{a}{}{,}
\usepackage{overpic}
\usepackage{fancyvrb}

\usepackage{multirow}

\begin{document}
\title{High-order aberration compensation with Multi-frame Blind Deconvolution and Phase Diversity image restoration techniques}

\author{G.B.\ Scharmer\inst{1,2}, M.G.\ L{\"o}fdahl\inst{1,2}, T.I.M.
  van Werkhoven\inst{1,3} \and J. de la Cruz Rodriguez\inst{1,2}}
\institute{Institute for Solar Physics, Royal Swedish Academy of
  Sciences, AlbaNova University Center, 106\,91 Stockholm, Sweden \and
  Stockholm Observatory, Dept. of Astronomy, Stockholm University,
  AlbaNova University Center, 106\,91 Stockholm, Sweden \and
  Sterrekundig Instituut Utrecht, Utrecht University, PO Box 80000,
  3508TA Utrecht, The Netherlands} \authorrunning{Scharmer,
  L{\"o}fdahl \& al.}

\titlerunning{High--order aberration compensation with MFBD and PD image restoration techniques}
\date{Final version: 7 Jul 2010}


\abstract
{For accurately measuring intensities and determining magnetic field
  strengths of small-scale solar (magnetic) structure, knowledge of
  and compensation for the point spread function is
  crucial. For images recorded with the Swedish 1-meter Solar Telescope (SST), restoration with Multi-Frame Blind Deconvolution (MFBD) and Joint Phase Diverse Speckle (JPDS) methods lead to remarkable improvements in image quality but granulation contrasts that are too low, indicating additional stray light. 
}
{We propose a method to compensate for stray light from high-order
  atmospheric aberrations not included in MFBD and JPDS processing.}
{To compensate for uncorrected aberrations, a reformulation of the
  image restoration process is proposed that allows the average effect
  of hundreds of high-order modes to be compensated for by relying on
  Kolmogorov statistics for these modes. The applicability of the
  method requires simultaneous measurements of Fried's parameter
  $r_0$. The method is tested with simulations as well as real data
  and extended to include compensation for conventional stray light.}
{We find that only part of the reduction of granulation
  contrast in SST images is due to uncompensated high-order
  aberrations. The remainder is still unaccounted for and attributed
  to stray light from the atmosphere, the telescope with its re-imaging
  system and to various high-altitude seeing effects.}
{We conclude that statistical compensation of high-order modes is a
  viable method to reduce the loss of contrast occurring when a
  limited number of aberrations is explicitly compensated for with
  MFBD and JPDS processing. We show that good such compensation is
  possible with only 10 recorded frames. The main limitation of the
  method is that already MFBD and JPDS processing introduces
  high-order compensation that, if not taken into account, can lead to
  over-compensation.}


\maketitle

\section{Introduction}
\label{sec:introduction}

Since the emergence of 3D simulations of solar convection, there has
remained a disturbing discrepancy between the measured contrast of
granulation and that obtained from simulations. It has been suspected
that the major reason for this discrepancy is atmospheric and
telescope stray light.  However, difficulties of accurately
characterizing such stray light, in particular as regards the far
wings of the corresponding point spread functions, have prevented firm
conclusions. Questions about the accuracy of the predicted intensity
contrasts obtained from simulations have also been raised and possible
effects of magnetic fields proposed as explanation of the reduced
contrast \citep{2007ApJ...668..586U}. Detailed comparisons of the
shapes of observed and simulated spectral lines, determined by
correlations between Doppler velocities and temperature variations,
however, show a remarkable agreement \citep{2009LRSP....6....2N}. In
addition, there is now excellent agreement between granulation RMS
contrasts obtained with independently developed 3D MHD codes
\citep{2009A&A...503..225W}.

Recent data obtained from Hinode show higher granulation contrasts
than obtained from most ground based solar telescopes, in spite of the
relatively modest aperture diameter of the solar optical telescope
(SOT) on Hinode \citep{2009A&A...503..225W}. The combined PSF of the
Broadband Filter Imager (BFI) and SOT, based on images recorded during
a Mercury transit and a solar eclipse and including stray light and
the effects of the large central obscuration and spider, was
determined by \cite{2008A&A...487..399W}. The obtained PSF leads to
reconstructed granulation contrasts that are remarkably close to those
of 3D simulations \citep{2009A&A...503..225W}. A simpler fit of the
PSF in terms of 4 Gaussians also leads to good agreements between
granulation contrast measured with BFI on Hinode and MHD simulations
\citep{2009A&A...501L..19M}.  In addition, agreement between simulated
images degraded to the resolution of the SOT spectro-polarimeter (SP)
shows good agreement, even without stray light correction
\citep{2008A&A...484L..17D}. This finally should settle any remaining
controversies about the RMS granulation contrasts obtained from 3D
solar convection simulations.

In this paper, we initiate a search for the origin of the ``missing''
granulation contrast in images observed with the Swedish 1-meter Solar
Telescope (SST) and restored with methods based on Multi-Frame Blind
Deconvolution \cite[MFBD; ][]{lofdahl02multi-frame}, such as Joint
Phase Diverse Speckle \citep[JPDS; ][]{paxman92joint} and Multi-Object
MFBD \cite[MOMFBD; ][]{noort05solar}. We have previously observed the
effects of truncating the wavefront expansion on contrasts and power spectra when comparing data restored with MFBD based methods to data
restored with Speckle interferometry \citep{paxman96evaluation,rouppe04penumbral}, and this observation is
the basis of the proposed investigation.

An important motivation for this study is Stokes measurements:
Magnetic structures inside and outside sunspots have much more
fine-structure than granulation. High spatial resolution and low
stray light is required for accurate field strength measurements of
such structures.  Present inversion techniques, such as Helix
\citep{2004A&A...414.1109L}, allow pixel-to-pixel compensation for
stray light through a parameter determined by the fits to the observed
Stokes profiles. It is however obvious that a stray light PSF cannot
vary significantly from one pixel to another and that the preferred
approach is compensation for stray light in pre-processing and to
carry out the inversions without allowance for stray light.

We note that polarization signals from isolated small-scale structures
recorded with high spatial resolution and high signal-to-noise (S/N)
show a halo of polarized light, probably originating from uncorrected aberrations and/or
stray light. Such isolated magnetic structures provide useful ``point sources'' that can be
used to validate measured PSFs in addition to measurements of
granulation contrast at different wavelengths.

The paper is organized as follows: We describe the MFBD/JPDS imaging
model and reconstruction in Sect.\@~\ref{sec:mfbd-imaging-model} and
propose a method for compensation of the PSF for high-order modes in
Sect.\@~\ref{sec:comp-uncorr-high}. We describe simulations and tests
made to validate the proposed method in Sect.\@~\ref{sec:simulations}
and apply the method to SST data in Sect.\@~\ref{sec:tests-with-real}.
Finally, we extend the method to include conventional stray light in
Sect.\@~\ref{sec:strayl-corr} and summarize the results in
Sect.\@~\ref{sec:conclusions}.

\section{MFBD imaging model and image reconstruction}
\label{sec:mfbd-imaging-model}

To allow compensation for aberrations not accounted for in processing
with MFBD based methods, we need to briefly review the imaging model
and image reconstruction process used. Although the MFBD based methods
include sophisticated schemes involving objects observed at different
wavelengths (MOMFBD) and with known aberration differences (JPDS), the
basic imaging model is simple and the reconstruction of the image well
defined.

In the MOMFBD implementation of \cite{noort05solar} the imaging process is modeled as a space-invariant (valid for
sufficiently small sub-fields) convolution between an unknown object
$f$ and a point spread function $t_k$, the Fourier transforms of
which are $F$ and $T_k$, where the index $k$ corresponds to a particular exposure. An additive Gaussian noise term $n_k$ is
assumed\footnote{A maximum-likelihood estimate with a Poisson noise model \citep{paxman92joint} would be more appropriate when photon noise is the dominating source of modelling error. The optimization process for the Poisson noise model is computationally much more demanding than for the additive Gaussian noise model.}. The Fourier transform $D_k$ of the observed image $d_k$ is
then related to $F$, $T_k$ and $N_k$ via
\begin{equation}
  D_k = F T_k + N_k .
  \label{eq:1}
\end{equation}
The assumption of additive Gaussian noise leads to a maximum-likelihood estimate of the object and transfer functions, equivalent to the solution of a conventional non-linear least-squares fit problem, and corresponds to the minimization of the scalar quantity $L$,
\begin{equation}
  L = \sum\limits_{u,v} \sum\limits_{k=1}^{K} |D_k - {\hat F} {\hat T_k}|^2 ,
  \label{eq:1b}
\end{equation}
where $u,v$ represents pixels in Fourier space. This equation directly
leads to an optimum estimate ${\hat F}$ of the true object $F$,
expressed in terms of the $K$ \emph{estimated} transfer functions
${\hat T}_k$ and associated observed images $d_k$
\begin{equation}
  {\hat F} = \sum\limits_{k=1}^{K} D_k {\hat T}_k^*\biggm/\sum\limits_{k=1}^{K}{|{\hat T}_k|^2},
  \label{eq:2}
\end{equation}
\citep{paxman92joint}, where $|{\hat T}_k|^2 = {\hat T}_k {\hat T}_k^*$
and ${\hat T}_k^*$ is the complex conjugate of ${\hat T}_k$. Note that
Eq.\@~(\ref{eq:2}) can be written as
\begin{equation}
  {\hat F} =\sum\limits_{k=1}^{K} w_k D_k/{\hat T}_k
  \label{eq:4}
\end{equation}
where $ D_k/{\hat T}_k$ are estimates of the object $F$ based on
\emph{single} frames and $w_k$ are weights, given by
\begin{equation}
  w_k = |{\hat T}_k|^2 \biggm/ \sum\limits_{k=1}^{K} |{\hat T}_k|^2.
  \label{eq:5}
\end{equation}
and $\sum w_k = 1$. This emphasizes that multi-frame deconvolution
assigns weights given by the absolute squares of the estimated
transfer functions associated with each image and at each spatial
frequency. This is the optimum way of combining measurements for which
$S/N$ varies because the transfer functions vary from one frame to
another while the RMS noise is the same for all measurements. In
particular, zero weight is given to spatial frequencies for which the
estimated transfer function pass through zero for an individual frame,
thus avoiding division by zero as may happen when $D_k$ is directly
divided by $\hat T_k$ to obtain a restored object. Eq.\@~(\ref{eq:4})
also means that a good image $D_k$, corresponding to an overall large
$|\hat T_k|$, is given (much) higher weight than a poor image in the
reconstruction of the object $F$. This weighting introduces a bias in
the image reconstruction that is of importance when compensating the
images for uncorrected high-order aberrations.

\section{Compensation for uncorrected high-order aberrations}
\label{sec:comp-uncorr-high}

MFBD and related image reconstruction techniques do not rely on statistical properties of seeing,
thereby allowing compensation for individual telescope aberrations and the
effects of an adaptive optics (AO) system in addition to seeing-induced aberrations. By estimating the transfer
function for each recorded frame, a small number of exposures can be
used to restore the object. However, the number of aberration
coefficients that can be determined from focused and defocused images
is necessarily limited. In less than perfect seeing, this leaves a
partially compensated PSF with enhanced wings. This corresponds to the
well-known ``halo'' seen in images of point-like objects recorded with
night-time telescopes and partial AO compensation of seeing
\citep[e.g., ][]{1992ESOC...42..475C}.

\subsection{Seeing statistics and residual aberrations expected}
\label{sec:seeing-stat-resid}

To appreciate the importance of residual high-order aberrations, we
note that \cite{2008A&A...484L..17D} found that aberrations as small
as 0.044~waves, corresponding to nearly perfect optics with a Strehl
ratio of 93\%, reduces the measured RMS contrast of granulation from
8.4\% to 8.1\% with Hinode's spectro-polarimeter. Thus, within the
framework of the present investigation, we clearly should consider any
residual aberrations that decrease the Strehl ratio below 90\%.

The effect of partially corrected aberrations on the Strehl ratio is
well-known from the work of Fried, Noll, Wang, Markey and others. We
refer to \citet{roddier99theoretical} for an overview of relevant
results. The residual wavefront variance for zonal correction can be
estimated as
\begin{equation}
  \sigma^2 = 0.34\, (D/r_0)^{5/3} N^{-5/6},
  \label{eq:9}
\end{equation}
where $\sigma$ is in radians, $D$ is the telescope diameter, $r_0$
is Fried's parameter, and $N$ is the number of independently corrected
aberration parameters. \emph{Optimum} compensation with a given number
of degrees of freedom is obtained with Karhunen--Lo\`eve (KL)
modes\footnote{This is however not true for MFBD/JPDS restoration of
  images obtained with AO.}. The residual wavefront variance,
estimated from a fit to data plotted in Fig.\@~3.1 of
\citet{roddier99theoretical}, is approximately
\begin{equation}
  \sigma^2 = 0.3\, (D/r_0)^{5/3} N^{-0.92} .
  \label{eq:15}
\end{equation}
The Strehl ratio $R$ can be estimated as
\begin{equation}
  R = \exp(-\sigma^2).
  \label{eq:10}
\end{equation}
We shall in the following estimates assume perfect zonal correction,
which in the case of AO correction would represent an unrealistically
high efficiency. For the SST, with a telescope diameter of 0.98~m, we
find that $R=0.81$ with $N=36$ aberration parameters when $r_0=22$~cm
and that $R>0.9$ only when $r_0>33$~cm. With good seeing corresponding
to $r_0=10$~cm (1\arcsec{} seeing) and very good seeing to $r_0=14$~cm
(0\farcs7 seeing), we need to compensate about 400 respectively 200
modes to reach a Strehl ratio of about 0.9. When $r_0=20$~cm (0\farcs5
seeing), we need to correct nearly 100 modes. Thus even with excellent
seeing, we should be concerned about the effects of uncorrected
aberrations in MFBD processing. These estimates also imply that the
problem of compensation of high-order aberrations in solar imaging is
primarily related to near-ground seeing, associated with a large
isoplanatic patch, since $r_0$ is typically more than a factor 3
smaller for the ground layer than for the high-altitude layers during day-time.

An important limitation at short wavelengths is that $r_0$ scales as
${\lambda}^{6/5}$. When $r_0=20$~cm at $\lambda = 630$~nm, $r_0$ is
only 11~cm at 390~nm. In this case we need to compensate nearly 100
aberrations at 630~nm and over 300 aberrations at 390~nm to reach a
Strehl ratio of 0.9. Even in excellent seeing, images recorded and
processed with conventional MFBD at 390~nm will be far from perfectly
corrected. It is therefore not surprising that measured RMS contrasts
of granulation with the SST show much larger discrepancies with 3D
simulations at short wavelengths than at long wavelengths.

\subsection{Compensation for high-order aberrations}
\label{sec:uncorr-high-order}

In MOMFBD processing of SST data, particularly from the CRISP
instrument \citep{scharmer08crisp}, a data set corresponding to a line
scan can consist of $\sim$1000 exposures in several cameras. However,
an estimated object from a particular wavelength and polarization
state is typically based on deconvolution of a relatively small
number of images ($\sim$10), degraded by residual low-order
aberrations from partial correction with a 37-electrode AO system and
uncorrected high-order atmospheric aberrations. As long as MFBD
processing compensates the residual aberrations partially corrected by
the AO system, we are to some extent justified in ignoring the
corrections made by the AO system: It is the residual aberrations
\emph{after} that AO correction that define the S/N of individual
images and the weighting implied by Eq.\@~(\ref{eq:4}).

Our goal is to estimate the effect on the transfer function of
uncompensated high-order aberrations, defined in Eq.\@~(\ref{eq:8}).
To do this, we will assume that these modes have amplitudes given by
the assumption of Kolmogorov turbulence. This is the same basic
assumption as used in Speckle Interferometry, but with the difference
that low-order modes are compensated individually and for each exposed
frame with MFBD. We conjecture that:
\begin{itemize}
\item there will be rather small variations of the PSF from one frame to another because residual wavefront errors, after correction on the order
  of 30 modes, depend on the accumulative effect of hundreds of modes
  rather than a few large-amplitude low-order modes.
\item a relatively small number of frames is therefore needed to
  obtain stable averages.
\end{itemize}

Solar speckle techniques rely on an average theoretical transfer function
obtained from Kolmogorov turbulence statistics and based on a large
number of recorded frames. This corresponds to compensation, in an
average sense, for an infinite number of aberration coefficients.
Recent developments allow solar speckle processing also of AO corrected
images \citep{puschmann06speckle,2007PhDT.........6W} but (as far as
we know) present solar speckle techniques do not include compensation for
telescope aberrations.

\subsubsection{The proposed method}
\label{sec:proposed-method}

Our goal is to develop a method that combines the advantages of
MFBD/JPDS and speckle methods. The approach taken is to compensate
individual frames for low-order modes and to add a compensation for
the \emph{average} effect of hundreds of high-order aberrations from
atmospheric turbulent seeing.

This compensation can be implemented in various ways. We have chosen
the following simple approach that has the advantage of not involving
the observed images but only their corresponding transfer functions.
We note that information about the ``true'' (exact) transfer functions
and the true object are encoded in the observed images. Ignoring the
noise term in Eq.\@~(\ref{eq:1}) and combining with
Eq.\@~(\ref{eq:2}), we obtain a relation between the estimated object
${\hat F}$, the true object $F$, and the corresponding exact and
estimated transfer functions $T_k$ and ${\hat T}_k$
\begin{equation}
  {\hat F} = F \sum\limits_{k=1}^{K} T_k {\hat T}_k^* \biggm/
  \sum\limits_{k=1}^{K}{|{\hat T}_k|^2} 
  \label{eq:6}
\end{equation}
We note that this is a relation of the form
\begin{equation}
  {\hat F} = S F,
  \label{eq:7}
\end{equation}
i.e., in the form of a multiplication of the true object with a
transfer function,
\begin{equation}
  S = \sum\limits_{k=1}^{K} T_k {\hat T}_k^* \biggm/ \sum\limits_{k=1}^{K}{|{\hat T}_k|^2}
  = \sum\limits_{k=1}^{K} w_k \frac{T_k}{{\hat T}_k},
  \label{eq:8}
\end{equation}
that is a weighted average of the ratio of the true and estimated
transfer equations. This equation does not solve the problem unless
the exact transfer functions are known. However, we can
\emph{estimate} $S$ by including known statistical properties of
\emph{atmospheric high-order} aberrations while using MFBD or JPDS
estimates of \emph{low-order aberrations for each individual
  exposure}. We propose the following estimate of $S$ from a
combination of aberration parameters determined by MFBD/JPDS
processing and statistical properties of atmospheric seeing:
\begin{equation}
  \hat S = \sum\limits_{k=1}^{K} \biggl\langle T_k {\hat T}_k^* 
  \biggm/ \sum\limits_{n=1}^{K}{|{\hat T}_n|^2} \biggr\rangle
  = \sum\limits_{k=1}^{K} \langle T_k \rangle {\hat T}_k^*  
  \biggm/ \sum\limits_{k=1}^{K}{|{\hat T}_k|^2}
  \label{eq:11}
\end{equation}
where angular brackets, $\langle\ldots\rangle$, indicate an ensemble
average over many independent realizations \emph{for each $k$
  separately} and where $T_k$ contains the same \emph{low-order} aberrations, obtained with MFBD or JPDS processing, as ${\hat T}_k$.

To estimate the effect of high-order aberrations in $T_k$, we add
random \emph{higher-order} KL modes with amplitudes given by Kolmogorov
statistics and average the transfer equation over many realizations to
obtain stable averages. The proposed method requires measurements of
$r_0$ at the time of recording the data. The preferable method for
accurate measurements of $r_0$ is via data from an open-loop wavefront
sensor located before the adaptive mirror, as implemented at the SST  \citep{2010A&A...513A..25S}. It is also possible, as
done at the Dunn telescope diameter \citep{2004SPIE.5490..184M}, to
combine closed-loop wavefront sensor data with the control matrix and
output voltages to estimate $r_0$ and residual low-order aberrations.
The quality of such measurements are to some extent limited by time
delays and inaccuracies in the control matrix, but experience with
night-time AO systems clearly indicates that good PSF
compensation is indeed possible with such data
\citep{1997JOSAA..14.3057V}.

\section{Simulations}
\label{sec:simulations}

To investigate the feasibility of the proposed method for compensation
of missing high-order aberrations in MFBD and JPDS processing,
numerical calculations and simulations were made.

\subsection{Ideal compensation with point sources}
\label{sec:simulations-tests}

\begin{figure*}[!t]
  \centering
  \includegraphics[bb=46 632 556 787,width=\linewidth]{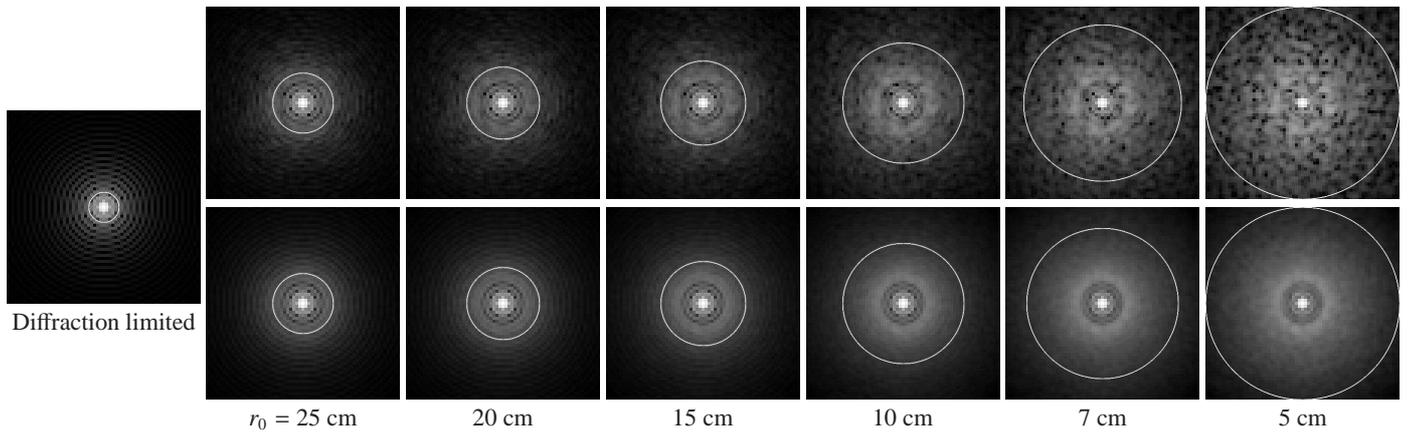}
  \caption{PSFs displayed in log scale. The circles mark 90\%
    encircled energy, see also Fig.\@~\ref{fig:energy}.  \textbf{Far
      left:} Diffraction limited. \textbf{Top:}~PSFs corresponding to
    $S$, i.e., true residual high-order aberrations for different
    $r_0$ as indicated.  
    \textbf{Bottom:}~Approximate PSFs, corresponding to $\hat S$,
    i.e., to the method proposed.
    The FOV shown is $4\farcs2\times 4\farcs2$ ($64\times 64$ pixels).
    The synthetic PSFs shown were calculated from 10 individual
    noise-free frames and with perfect compensation for the 36 lowest
    KL-modes.} 
  \label{fig:PSFs}
\end{figure*}

\begin{figure*}[!t]
  \centering
  \begin{minipage}[t]{0.32\linewidth}
    \includegraphics[bb=21 9 493 343,clip,width=\linewidth]{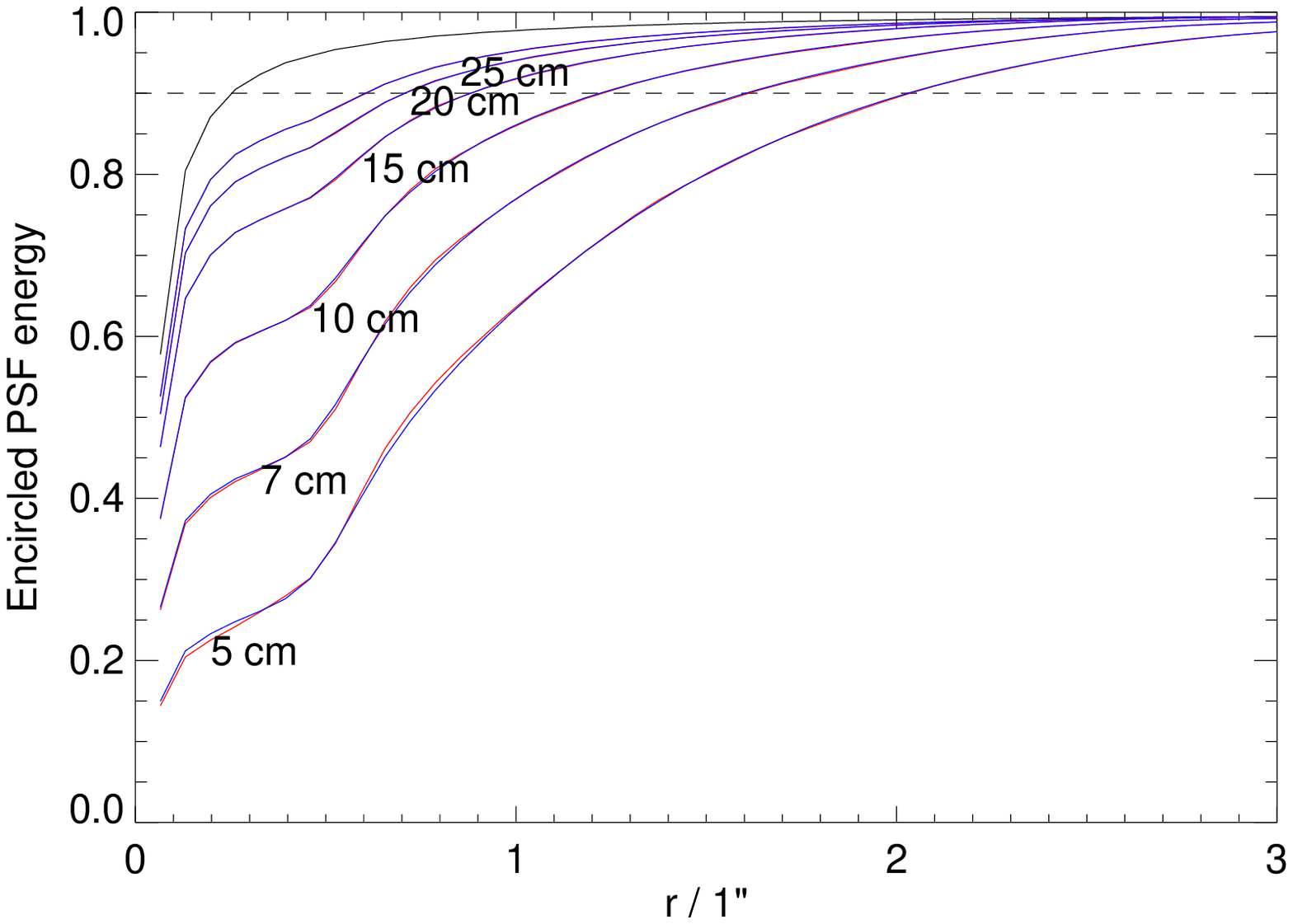}
    \caption{Encircled PSF energy for different $r_0$ as indicated in
      the figure. Red:~PSFs based on $S$; Blue:~PSFs based on $\hat
      S$; Black:~diffraction limited PSF; Black dashed: 90\% level.}
    \label{fig:energy}
  \end{minipage}
  \hfill
  \begin{minipage}[t]{0.32\linewidth}
    \includegraphics[bb=21 9 493 343,clip,width=\linewidth]{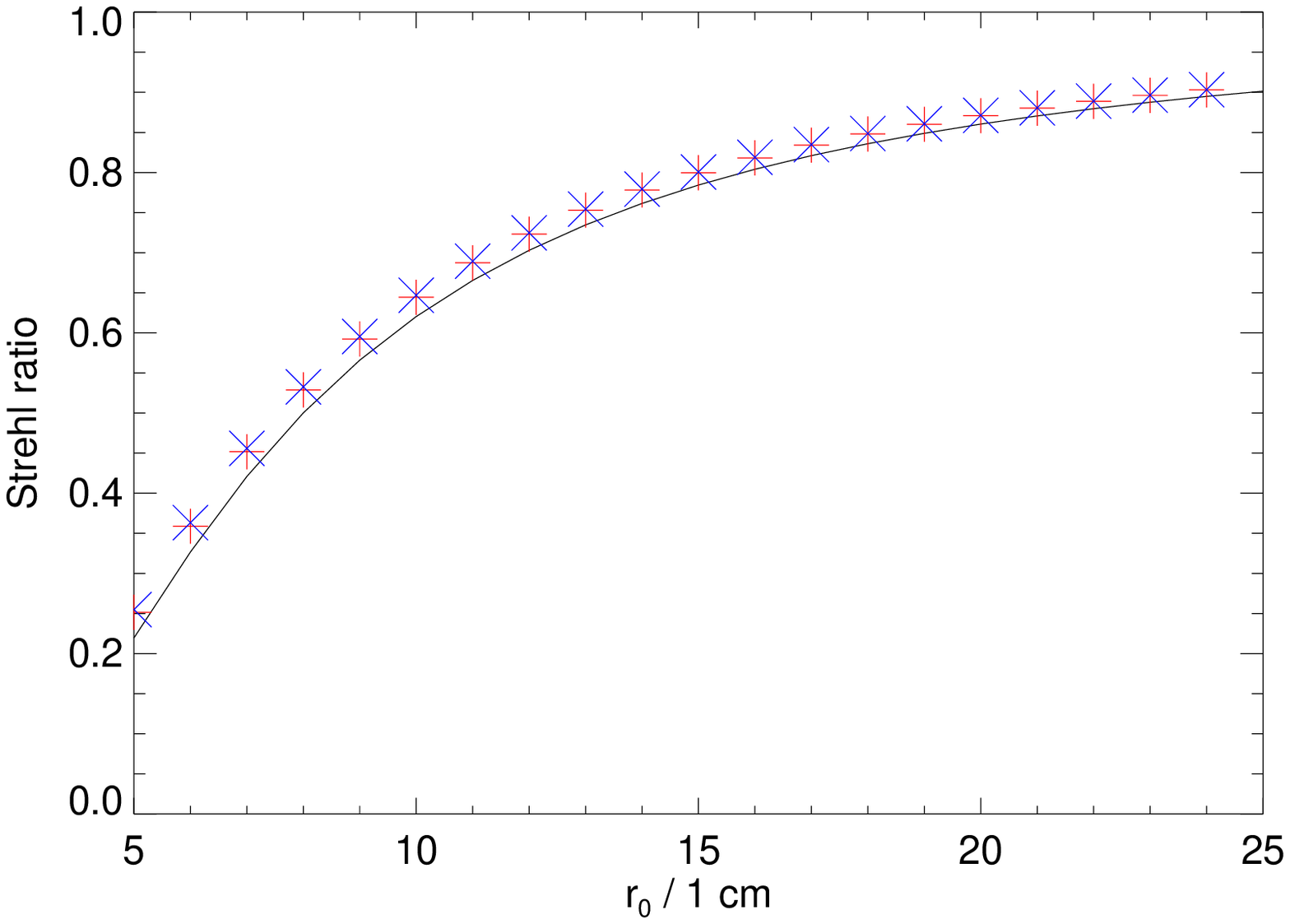}
    \caption{Strehl ratios as a function of $r_0$. Solid line:
      Eqs.\@~(\ref{eq:15}) and (\ref{eq:10}) with $N=37$; Red plus
      ($+$) symbols:~PSFs based on $S$; Blue cross ($\times$)
      symbols:~PSFs based on~$\hat S$.}
    \label{fig:strehl}
  \end{minipage}
  \hfill
  \begin{minipage}[t]{0.32\linewidth}
    \includegraphics[bb=21 9 493 343,clip,width=\linewidth]{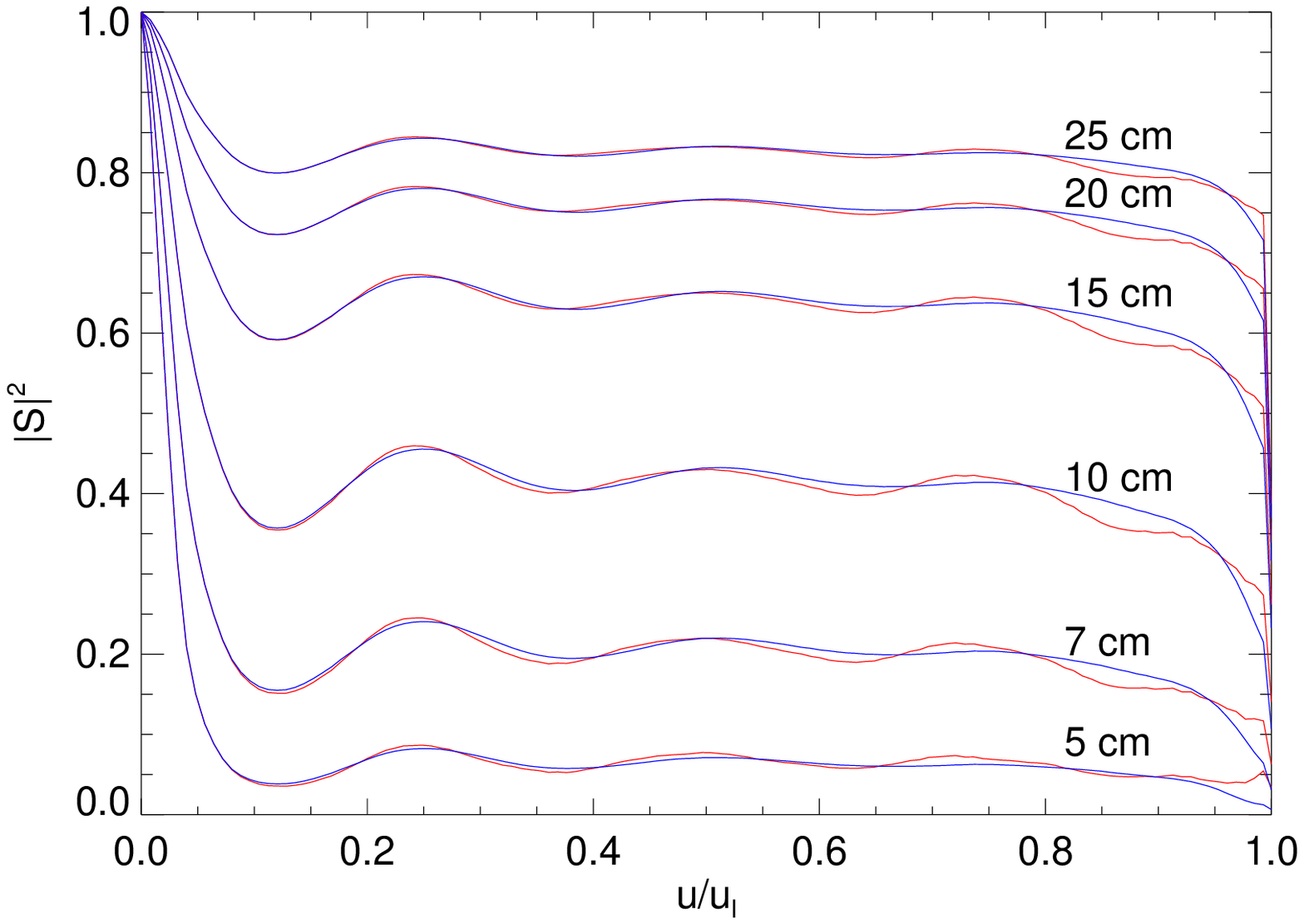}
    \caption{Power spectra (angular averages) of $S$ (red) and $\hat
      S$ (blue) for $r_0$ as indicated.} 
    \label{fig:power} 
  \end{minipage}
\end{figure*}
\begin{figure*}[!t]
  \centering
  \includegraphics[bb=46 620 556 787,width=\linewidth]{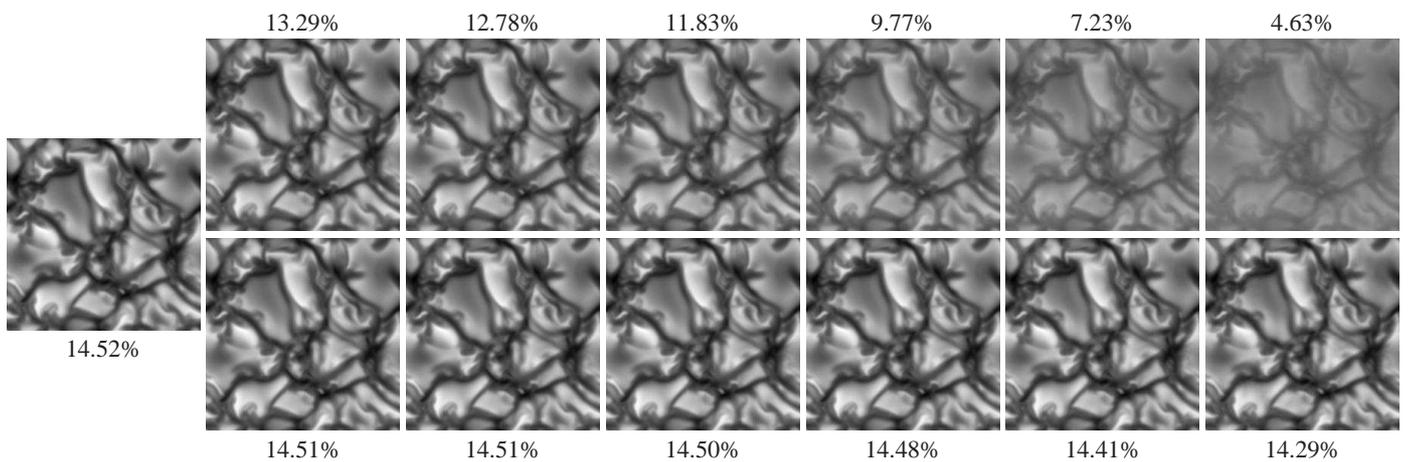}
  \caption{Synthetic images calculated at a wavelength of 630~nm.
    \textbf{Far left:}  Original image. \textbf{Top:}~Low-pass images
    degraded by high-order aberrations, i.e., by $S$, corresponding to
    $r_0=25$, 20, 15, 10, 7 and 5~cm, resp.\@ (same layout as
    Fig.\@~\ref{fig:PSFs}).
    \textbf{Bottom:}~Degraded images compensated by use of the method
    described, i.e., by $\hat S$.
    All images are scaled between min and max of the original image
    and low-pass filtered to 90\% of the SST diffraction limit. The
    numbers above and below the image tiles are the RMS contrasts in
    percent of the mean intensity
    ($100\times\text{RMS}/\text{mean}$).}
  \label{fig:images}
\end{figure*}

We used KL functions based directly on the theory of
\citet{fried78probability}, as implemented by \citet{dai95modal}.
These functions are orthogonal on a circular aperture and
statistically independent for Kolmogorov turbulence. For such
turbulence, the variances depend only on $r_0$. To produce random
wavefronts following Kolmogorov statistics, 1001 random numbers were
drawn from a standard normal distribution, scaled with the square root
of the theoretical variances and used as coefficients for KL functions
4--1004 (in decreasing variance order). The resulting wavefronts were
scaled to different values of Fried's parameter $r_0$. To simulate the
effect of partial correction with an efficient AO system, we reduced
the amplitudes of KL coefficients 4--37 with a factor~4 (this represents an overestimate of the actual efficiency). Piston
(coefficient 1) is ignored because it does not contribute to the OTFs
and tip-tilt correction (coefficients 2 \& 3) was assumed to be
perfect, corresponding to images recorded with sufficiently short exposure times to remove changes in image position and blurring during the exposure.

The residuals of the AO-corrected aberrations were used to represent
the estimate of the transfer equations $\hat T_k$. For this
calculation we assumed 10 observed images with independently obtained
wavefronts. We computed the ``corrective'' transfer function $S$ from
Eq.\@~(\ref{eq:8}) by using the actual high-order aberrations
corresponding to each $T_k$ and then the approximate version, $\hat
S$, using the statistical averages (based on 100 realizations of the
high-order tail) of the transfer functions, as defined in
Eq.\@~(\ref{eq:11}). We finally compared the exact and estimated
transfer functions $S$ and~$\hat S$. This corresponds to perfect MFBD
correction of the first 36 KL modes and perfect knowledge of $r_0$.

Figure~\ref{fig:PSFs} shows the PSFs corresponding to $S$ and $\hat
S$, scaled such that the wings of the PSFs can be seen, for values of
$r_0$ in the range 5--25~cm and $\lambda=630$~nm. Calculations were
made for a 98~cm telescope diameter, corresponding to the SST, with
critical sampling at $\lambda=630$~nm, corresponding to
$0\farcs066/\text{pixel}$. The field of view (FOV) shown is
$4\farcs2\times 4\farcs2$. As expected, the true PSFs show a speckled
structure in the wings, but smoothed by the averaging effect obtained
by combining ten images. The corresponding approximate PSFs show much
less structure in the wings but are otherwise similar to the actual
PSFs. It can also be seen that the effective diameter of the PSF,
defined as that containing 90\% of the energy of the PSF, increases
with decreasing $r_0$. When $r_0$ equals 25~cm, that diameter is
approximately 1\farcs1, when $r_0$ is 5~cm, it increases to
approximately~4\arcsec{}\footnote{As discussed in Sect. 5, the
  increasingly larger diameter of the PSF in degrading seeing
  restricts the use of very small subfields needed with MFBD/JPDS
  methods to deal with anisoplanatism. A similar restriction applies
  when the images are intentionally degraded by including a defocused
  imaging channel, used with JPDS processing.}.
Figure~\ref{fig:energy} shows the encircled energy as function of
radius for the approximate and exact PSFs. It is clear that the
encircled energy of the approximate PSF follows that of the actual PSF
nearly exactly. In Fig.\@~\ref{fig:strehl} we show calculated Strehl
ratios of the PSFs as function of $r_0$. The exact and approximate
PSFs give nearly exactly the same Strehl ratios and also agree well
with what is expected for perfect KL correction of 36 modes from
Eqs.\@~(\ref{eq:15}) and (\ref{eq:10}). Finally,
Fig.\@~\ref{fig:power} shows radially averaged power spectra for the
transfer functions corresponding to the exact and approximate PSFs,
again showing an excellent agreement between the two at all spatial frequencies.
We also refer the reader to Fig.\@~12 by \citet{rouppe04penumbral}, where similar effects of seeing on observed power spectra of penumbral fine structure are discussed.

\subsection{Ideal compensation with granulation images}
\label{sec:simulations-granulation}
To investigate the effects of uncompensated high-order aberrations on
granulation images, we used synthetic images calculated from a field-free 3D MHD
simulation \citep{1998ApJ...499..914S}; these simulation data were
kindly provided by Mats Carlsson. The synthetic images were calculated
at a wavelength of 630~nm and were degraded to a resolution
corresponding to 90\% of the diffraction limit of the SST while
keeping the image scale of the original synthetic images. The images
were then degraded by Fourier multiplication with $S$ based on 10
wavefronts. This corresponds to an MFBD estimate $\hat F$ based on 10
observed images. $\hat F$ was then corrected using an approximate
$\hat S$ based on the same low-order aberrations as $S$ and a
high-order compensation with 1000 KL-modes according to
Eq.\@~(\ref{eq:11}). The original, degraded and restored images are
shown in Fig.\@~\ref{fig:images} together with their RMS contrasts.
Note that the degraded images differ from the original image in
contrast but not in fine structure shown. The main effect of the
higher-order aberrations is to add stray light, decreasing the RMS
contrast, whereas all small-scale features of the original image are
retained and restored to full contrast with the approximate PSFs. As
shown in Fig.\@~\ref{fig:images}, the loss in contrast is from 14.5\%
to 11.8\% when $r_0$ is 15~cm, or a reduction of the RMS contrast by
nearly~20\%. This corresponds to very good seeing conditions. In more typical seeing conditions, when $r_0$ is 10~cm or smaller, the effects are larger.

We conclude that whereas details of the exact and approximate PSFs
certainly differ, the overall effect of the high-order aberrations is
to add spatial stray light to the images. The proposed method for
statistical compensation of high-order aberrations in principle should work
very well.

\subsection{Inversion tests with granulation images}
\label{sec:invers-test}

With perfect determination of the low-order modes, correction of the
high order modes in a \emph{statistical} sense gives excellent
results. However, we also need to investigate the effect of realistic
errors from the low-order modes estimated with MFBD and JPDS
techniques.

For this test, we calculated 1000 random wavefronts corresponding to
Kolmogorov statistics, with reduced amplitudes for the first 36 KL
coefficients as described above. These wavefronts, scaled to different
$r_0$, were used to construct sets of degraded granulation images
with and without phase diversity, corresponding to an added focus
shift, of 1 wave peak-to-valley. We used synthetic images calculated
from the 3D MHD code as true objects,
compressed by a factor 2 (by cropping the Fourier transform of the
image). The image scale was set to match that of SST/CRISP during the
2009 observing season, $0\farcs059/\text{pixel}$, corresponding to
about 12\% oversampling at 630~nm.

The degraded synthetic images were processed with the MOMFBD program
in various ways (MFBD or JPDS, different numbers of realizations,
different subfield sizes (256 pixels = 15\arcsec{}, 128 pixels =
7\farcs6, 80 pixels = 4\farcs7), different number of estimated
wavefront parameters, with and without added noise).

\begin{figure*}[!t]
  \centering
\subfloat[]{\includegraphics[bb=34 24 196 283,clip,width=0.16\linewidth]{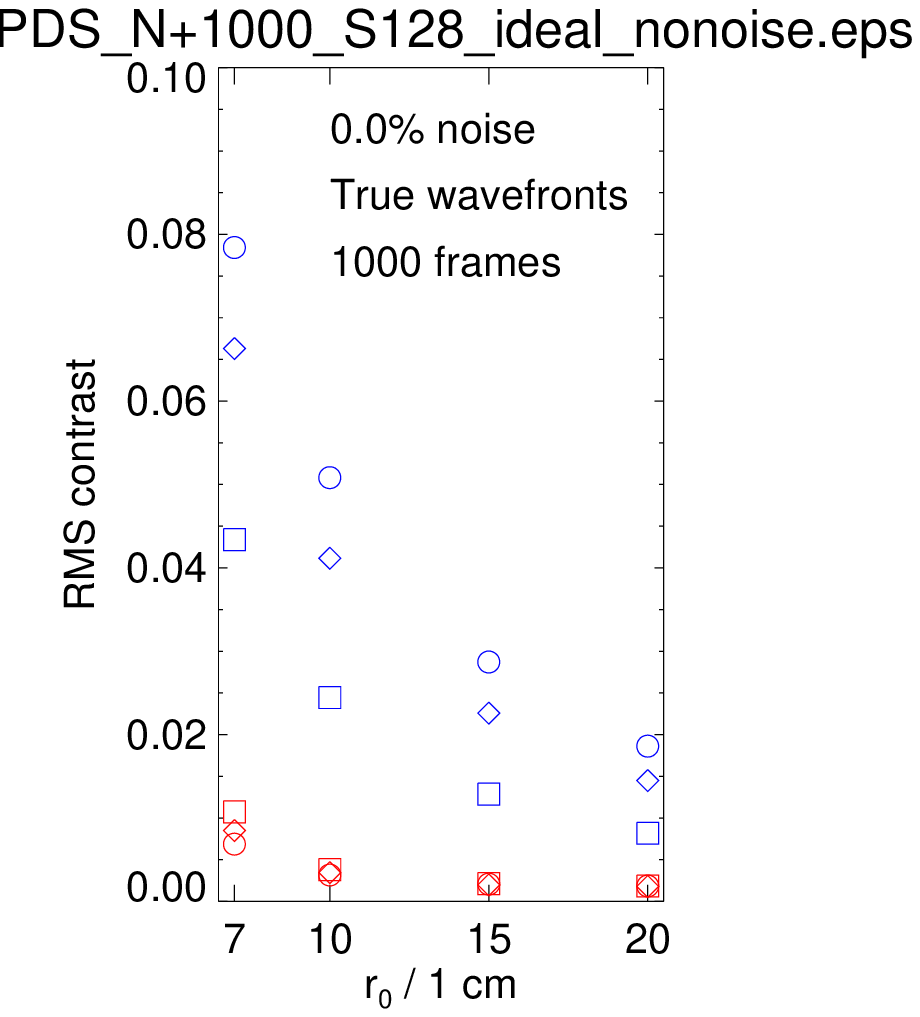}}
\subfloat[]{\includegraphics[bb=34 24 196 283,clip,width=0.16\linewidth]{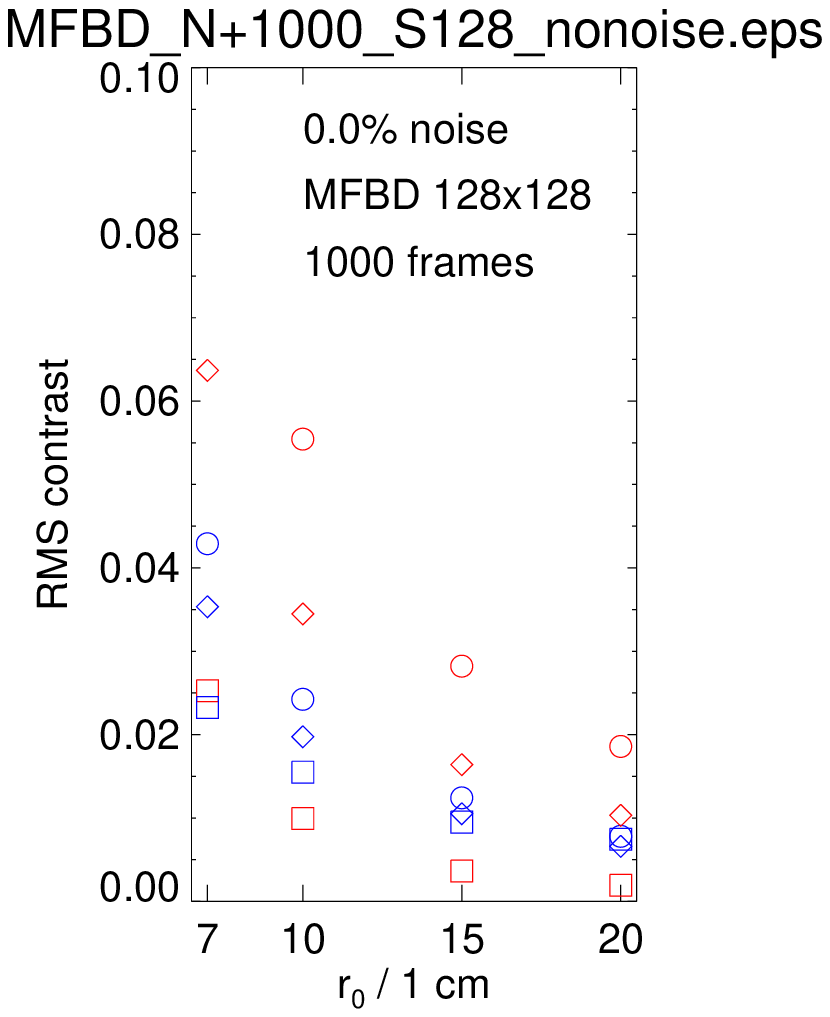}}
\subfloat[]{\includegraphics[bb=34 24 196 283,clip,width=0.16\linewidth]{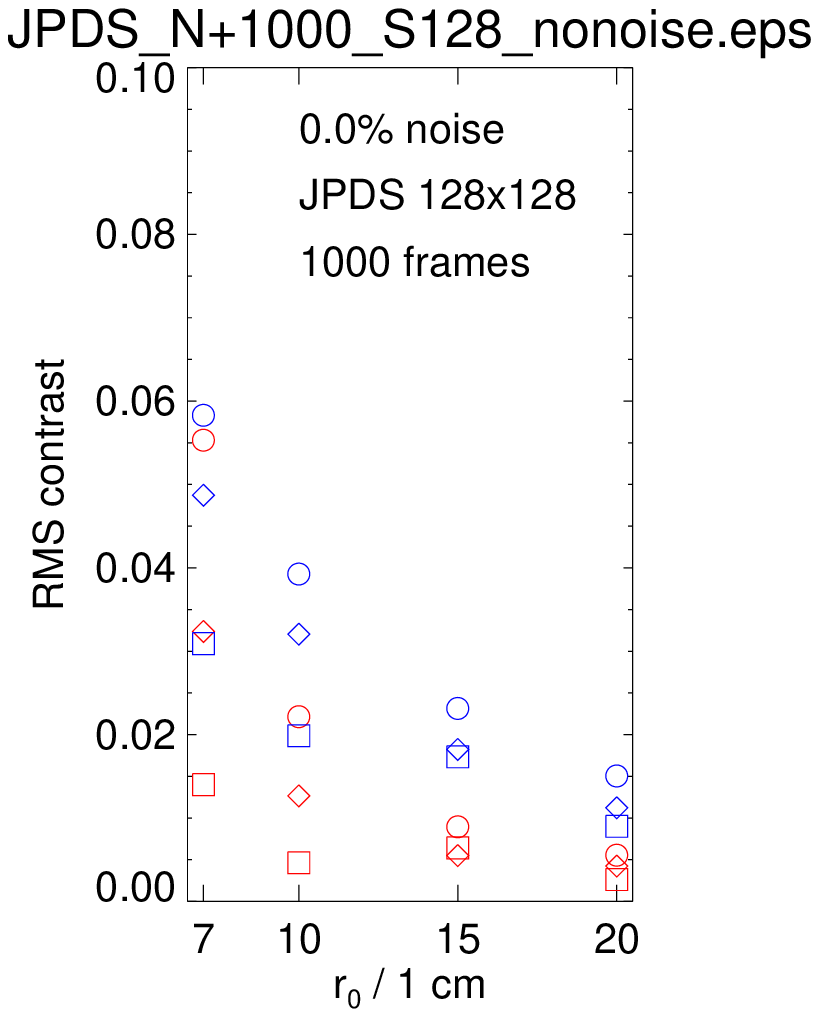}}
\subfloat[]{\includegraphics[bb=34 24 196 283,clip,width=0.16\linewidth]{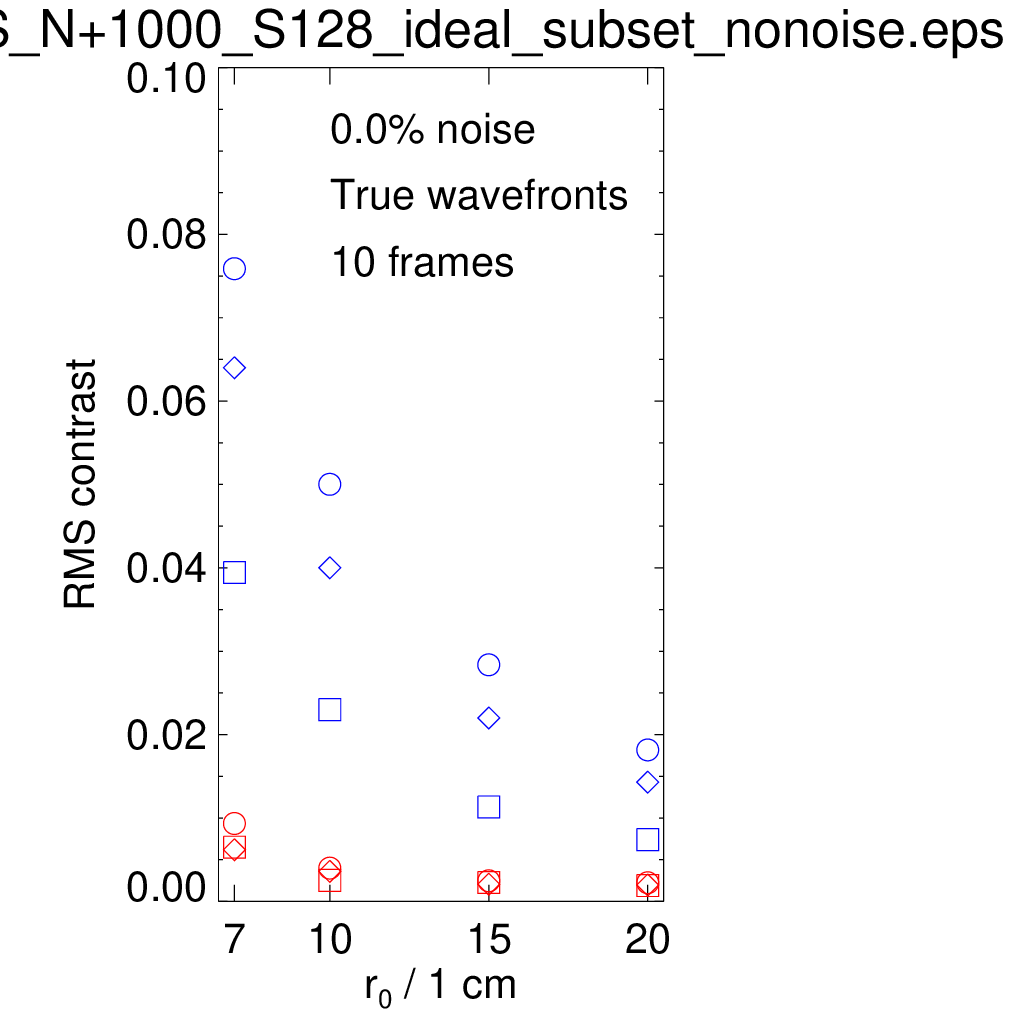}}
\subfloat[]{\includegraphics[bb=34 24 196 283,clip,width=0.16\linewidth]{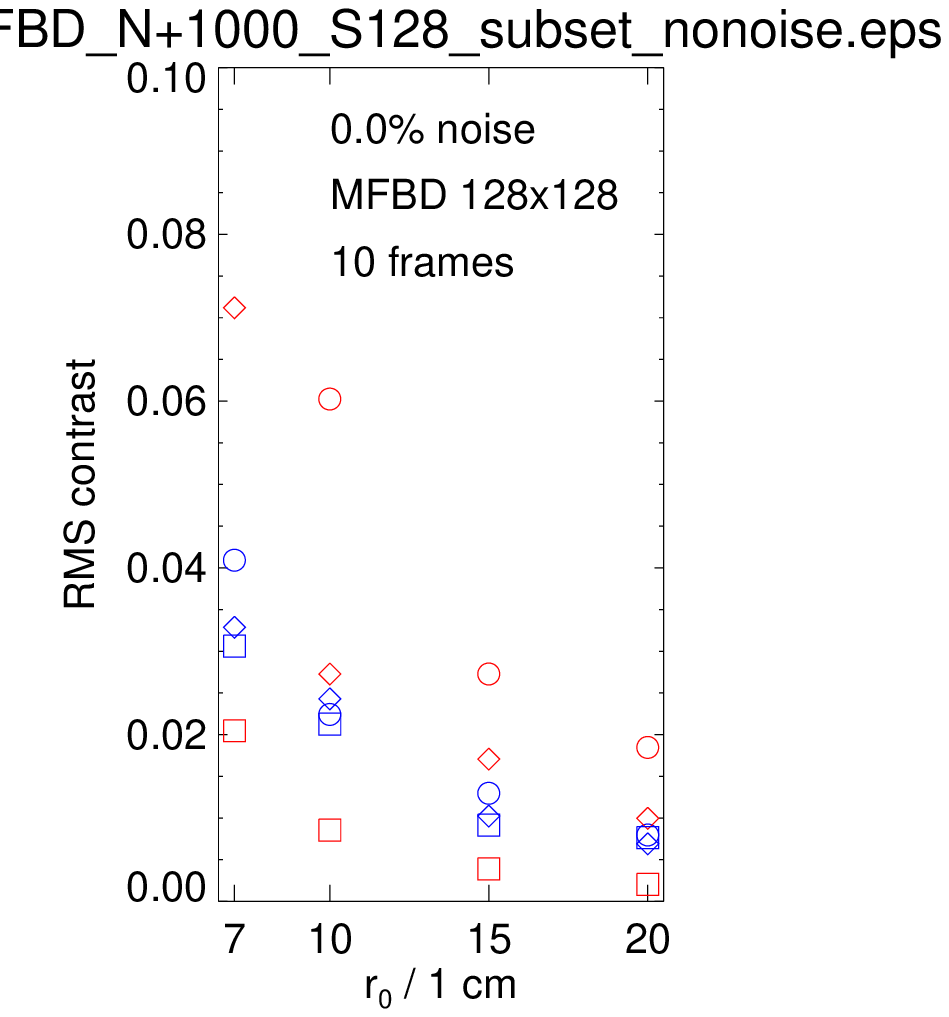}}
\subfloat[]{\includegraphics[bb=34 24 196 283,clip,width=0.16\linewidth]{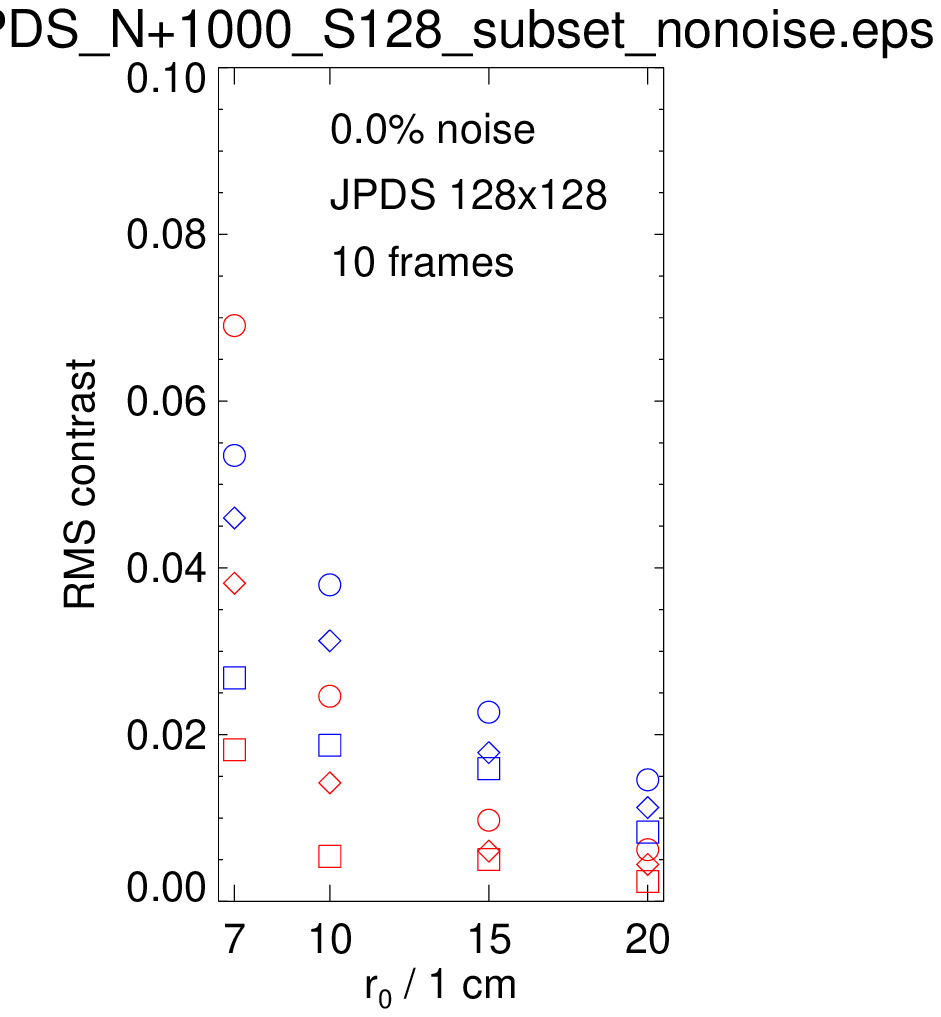}}
 \\
\subfloat[]{\includegraphics[bb=34 24 196 283,clip,width=0.16\linewidth]{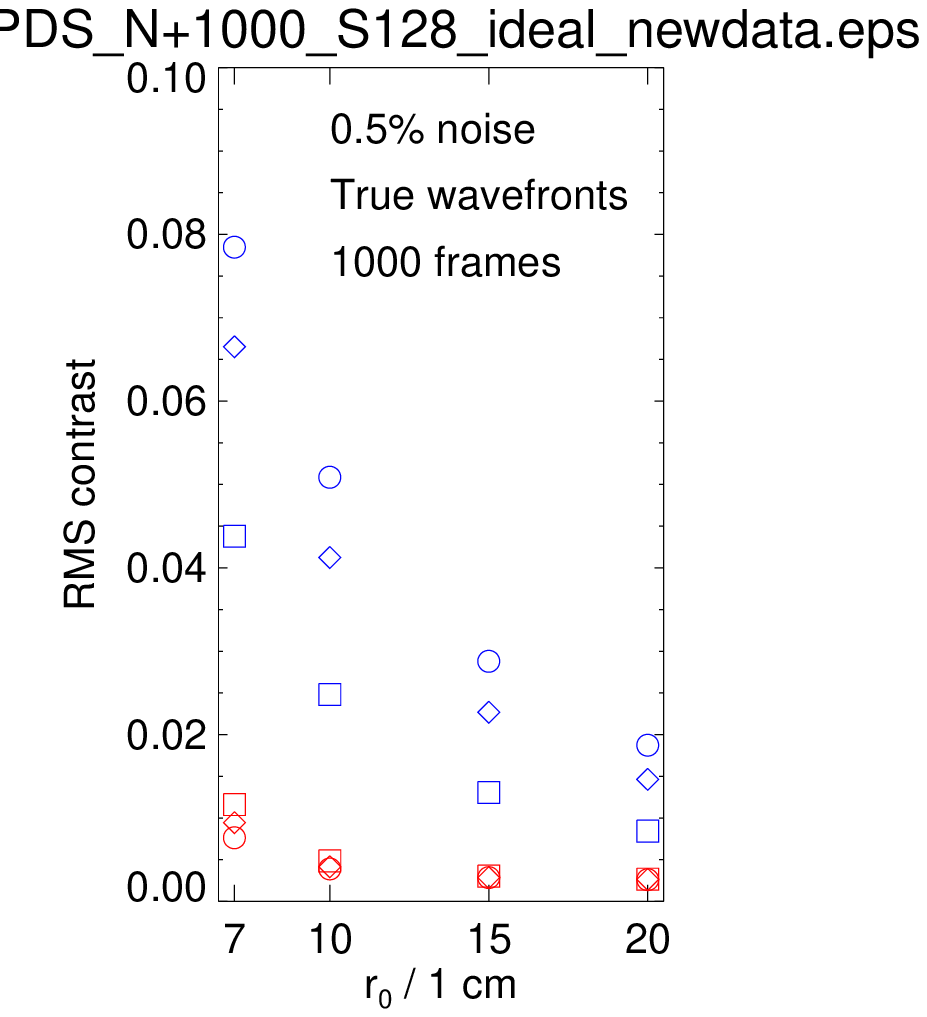}}
\subfloat[]{\includegraphics[bb=34 24 196 283,clip,width=0.16\linewidth]{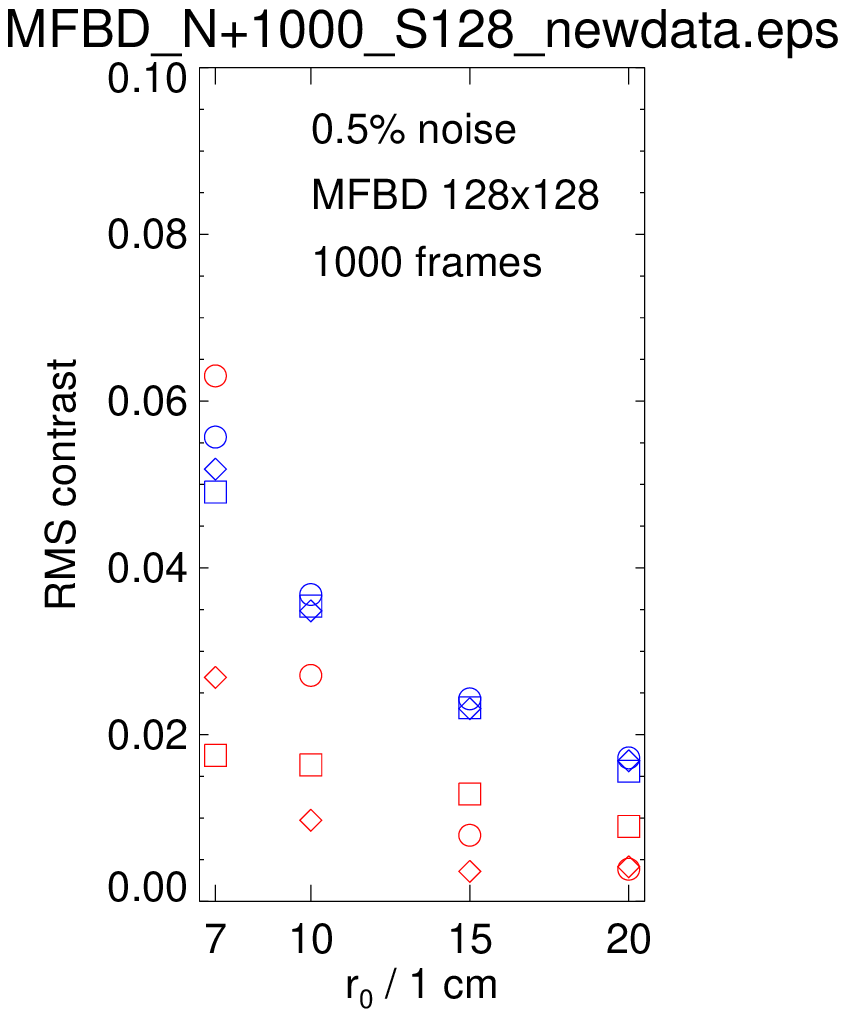}}
\subfloat[]{\includegraphics[bb=34 24 196 283,clip,width=0.16\linewidth]{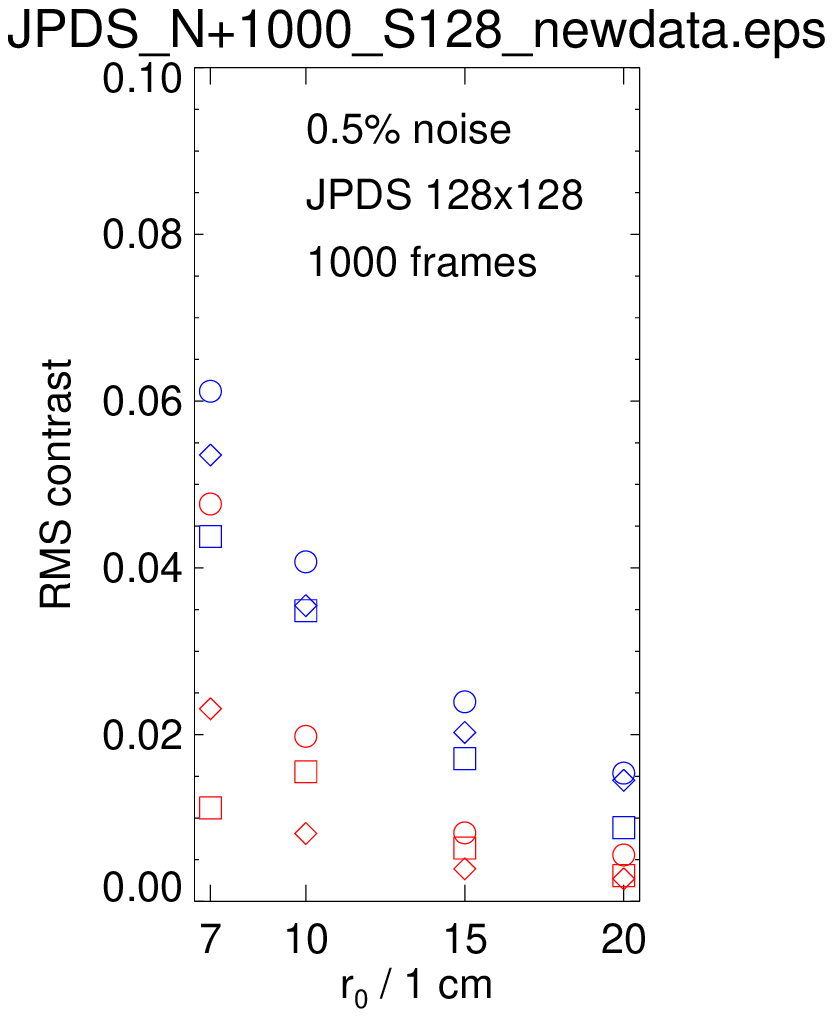}}
\subfloat[]{\includegraphics[bb=34 24 196 283,clip,width=0.16\linewidth]{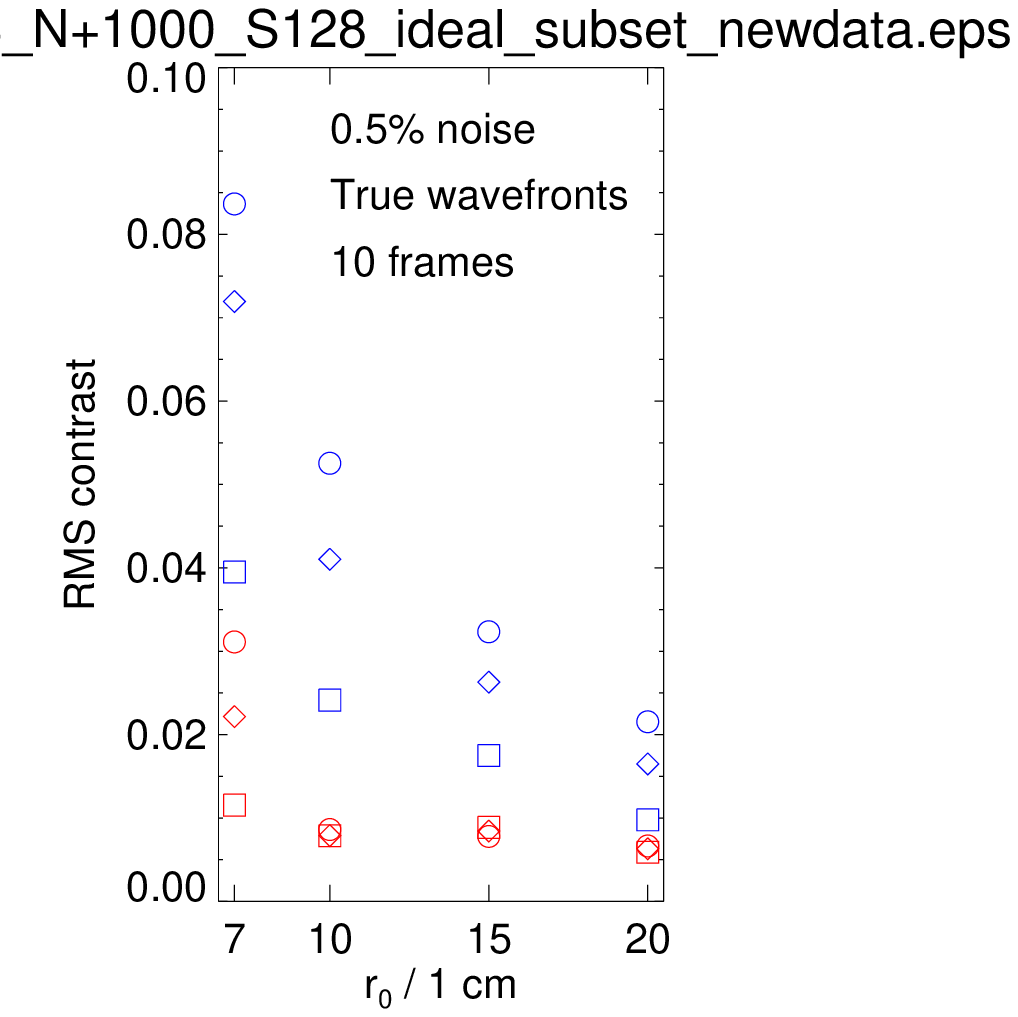}}
\subfloat[]{\includegraphics[bb=34 24 196 283,clip,width=0.16\linewidth]{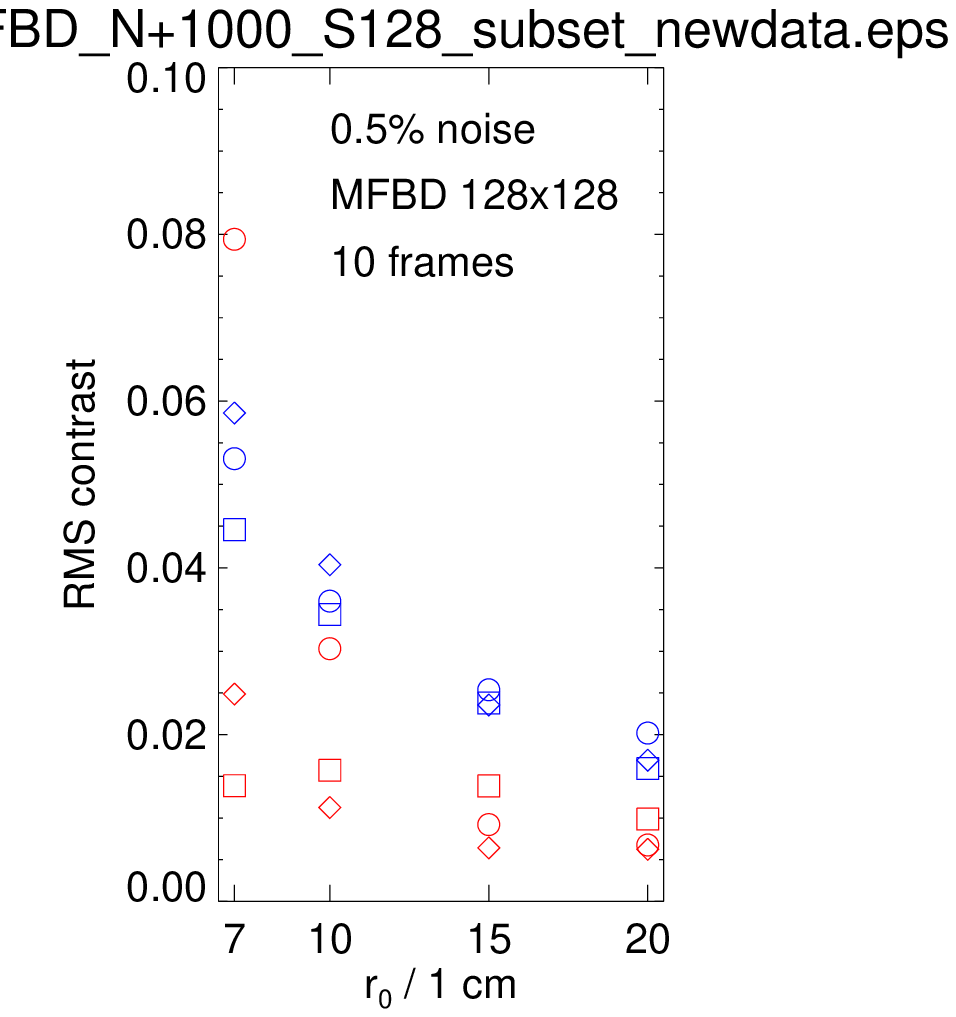}}
\subfloat[]{\includegraphics[bb=34 24 196 283,clip,width=0.16\linewidth]{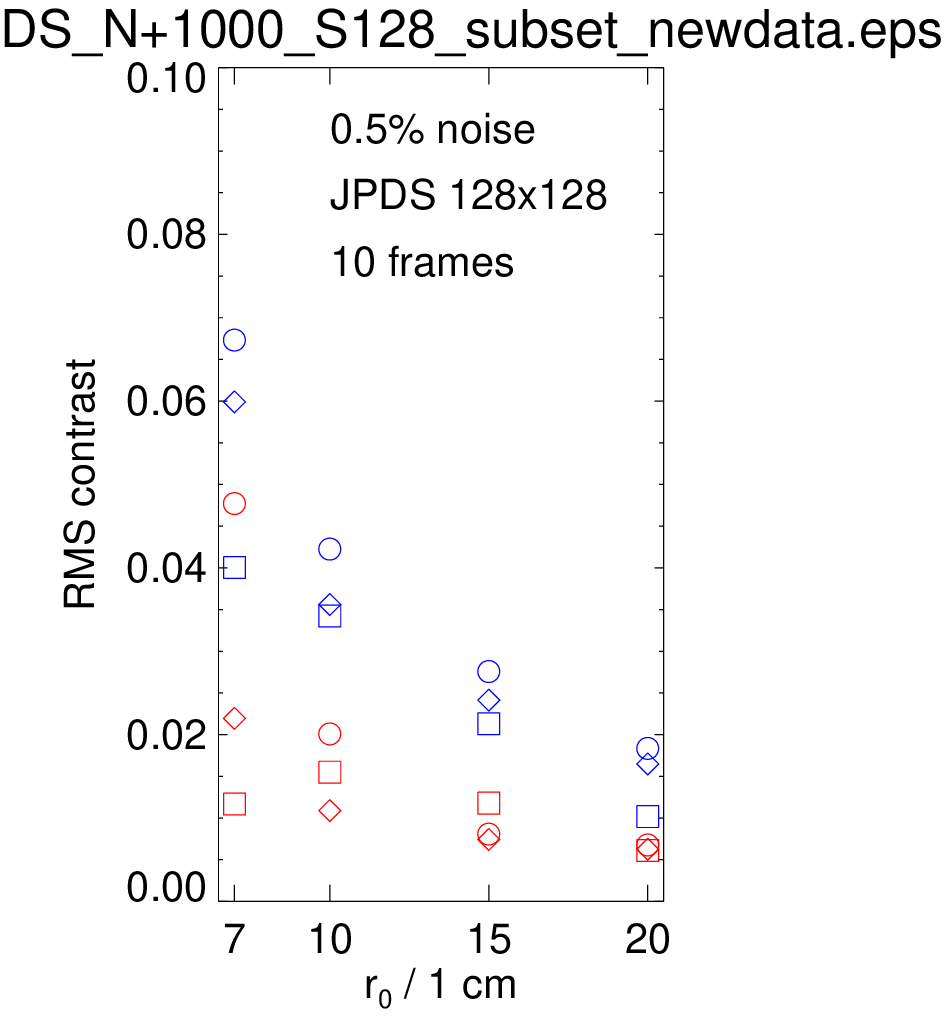}}
  \caption{RMS intensity error (in \% of the average intensity) of
    restored granulation images as a function of $r_0$ (in cm) after
    correcting $M$ modes. Circles ($\medcirc$):~$M=35$; Diamonds
    ($\Diamond$):~$M=50$; Squares ($\square$):~$M=100$.
    Blue:~Restored without $\hat S$ compensation;
    Red:~Restored with $\hat S$ compensation.}
  \label{fig:final}
\end{figure*}
\begin{figure*}[!t]
  \centering
  \subfloat[]{\includegraphics[bb=38 24 196 283,clip,width=0.16\linewidth]{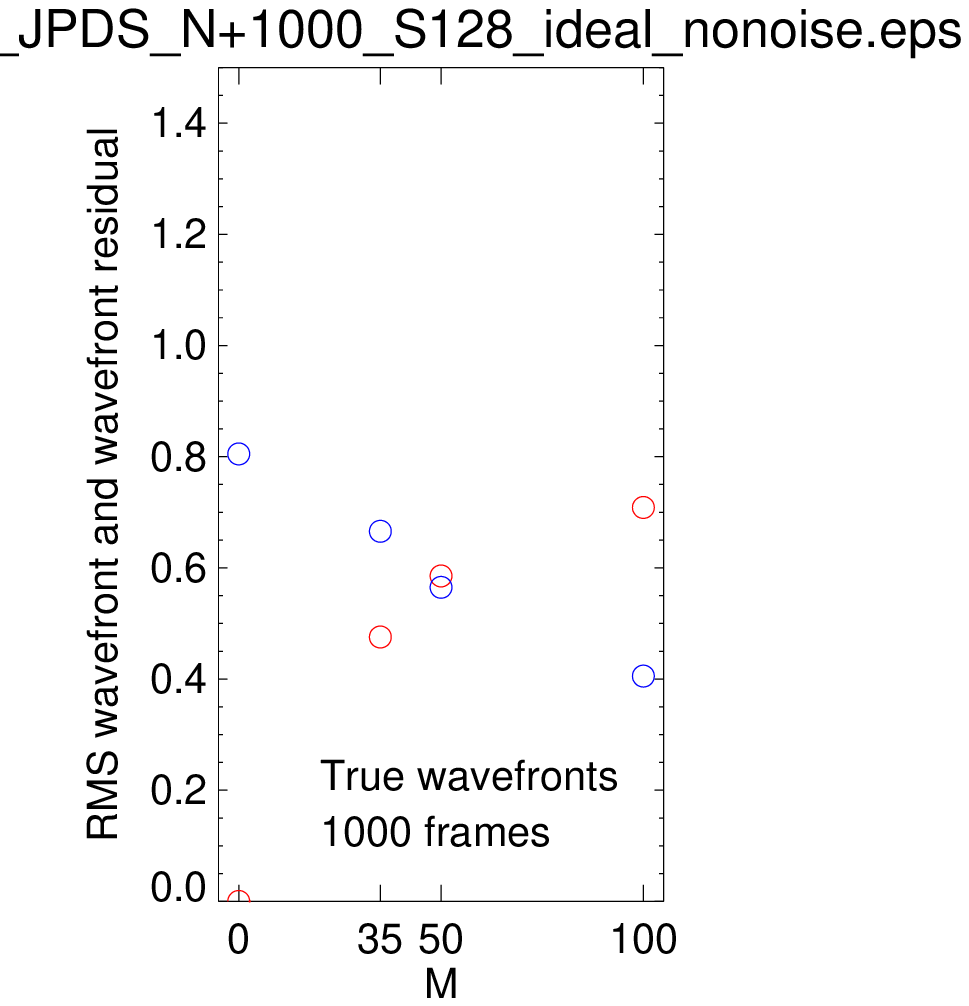}}
  \subfloat[]{\includegraphics[bb=38 24 196 283,clip,width=0.16\linewidth]{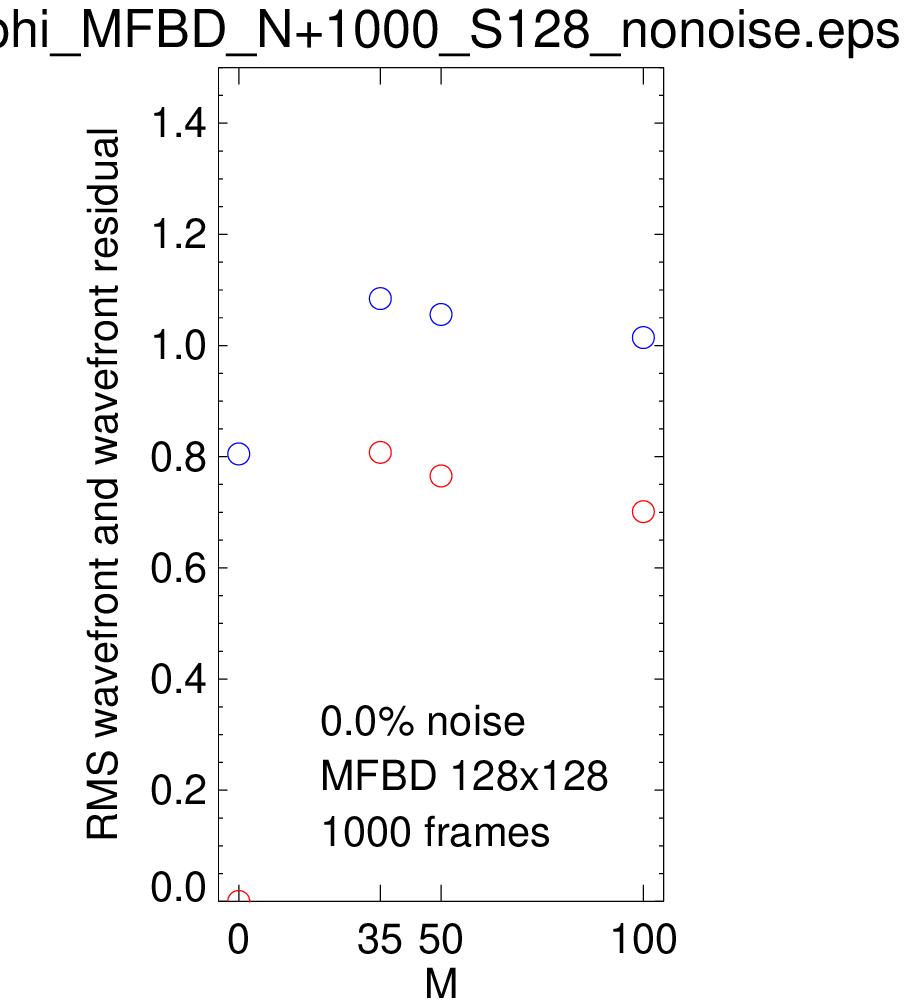}}
  \subfloat[]{\includegraphics[bb=38 24 196 283,clip,width=0.16\linewidth]{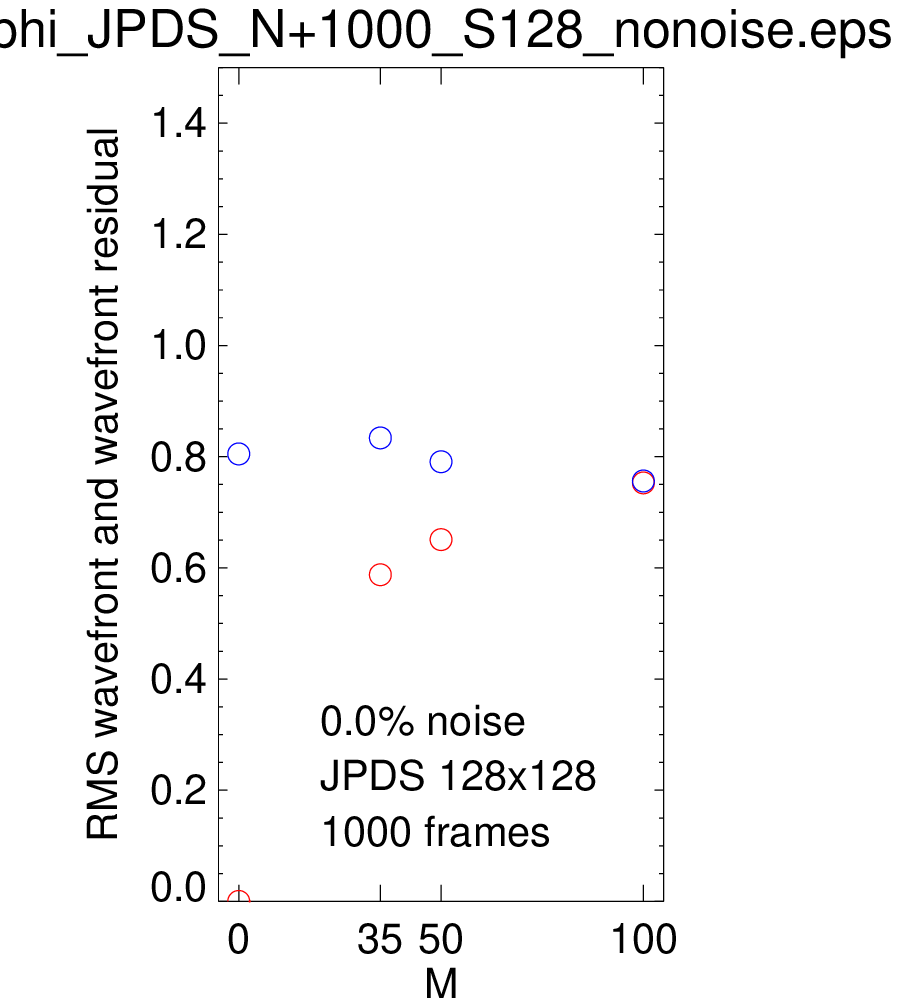}}
  \subfloat[]{\includegraphics[bb=38 24 196 283,clip,width=0.16\linewidth]{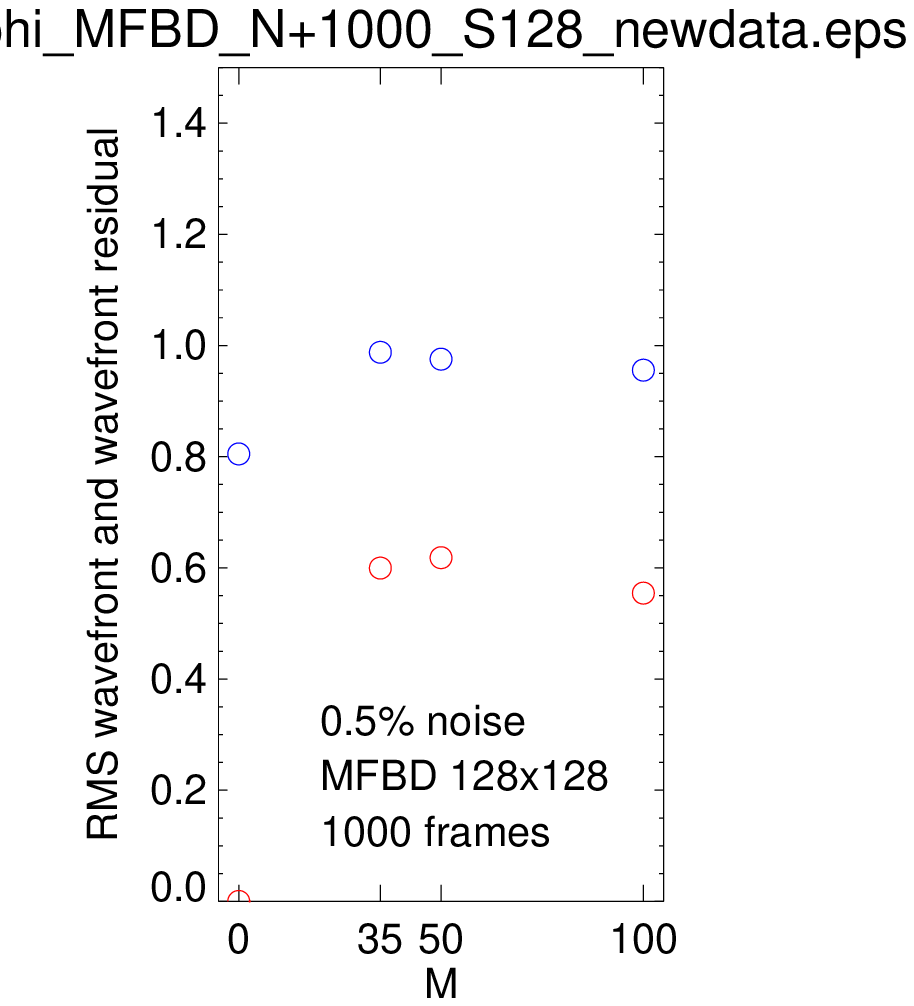}}
  \subfloat[]{\includegraphics[bb=38 24 196 283,clip,width=0.16\linewidth]{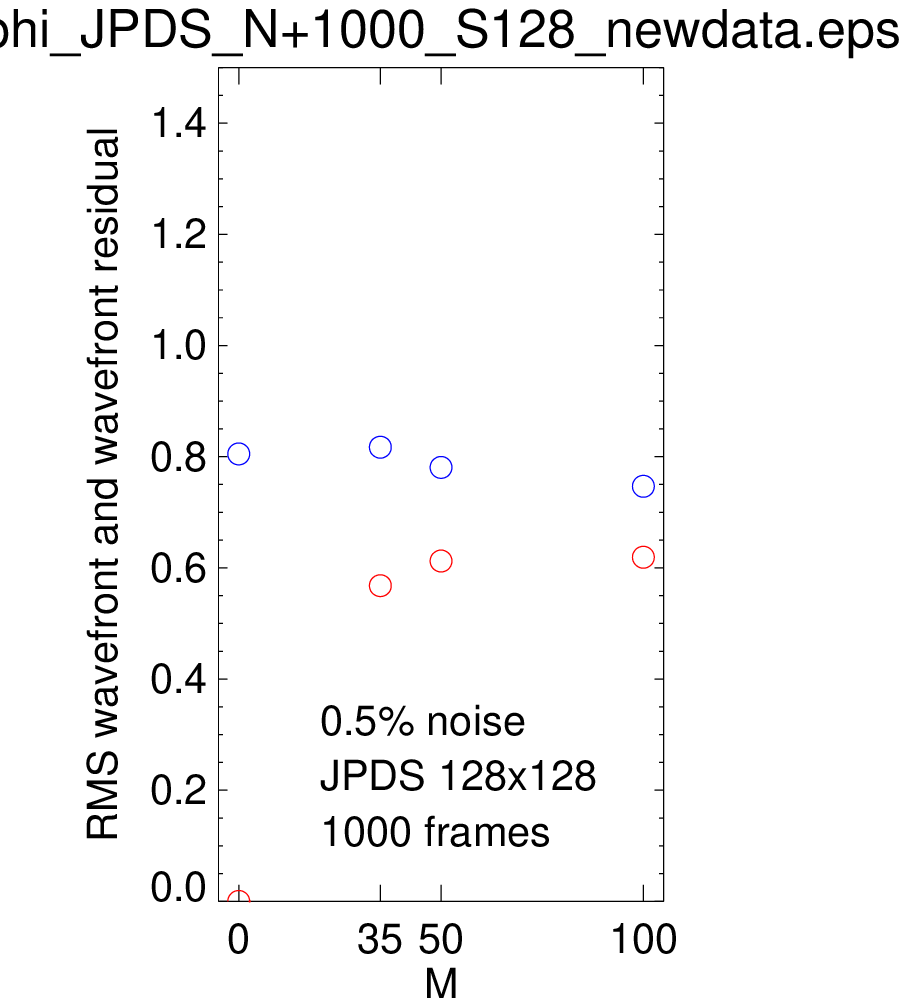}}
  \caption{Wavefront quantities as a function of $M$
    (number of corrected modes) for $r_0=10$~cm.
    Red:~RMS wavefront in rad;
    Blue:~RMS wavefront residual in rad.}
  \label{fig:final_phi}
\end{figure*}

\begin{figure*}[!t]
  \centering
  \subfloat[]{\includegraphics[bb=38 24 196 283,clip,width=0.16\linewidth]{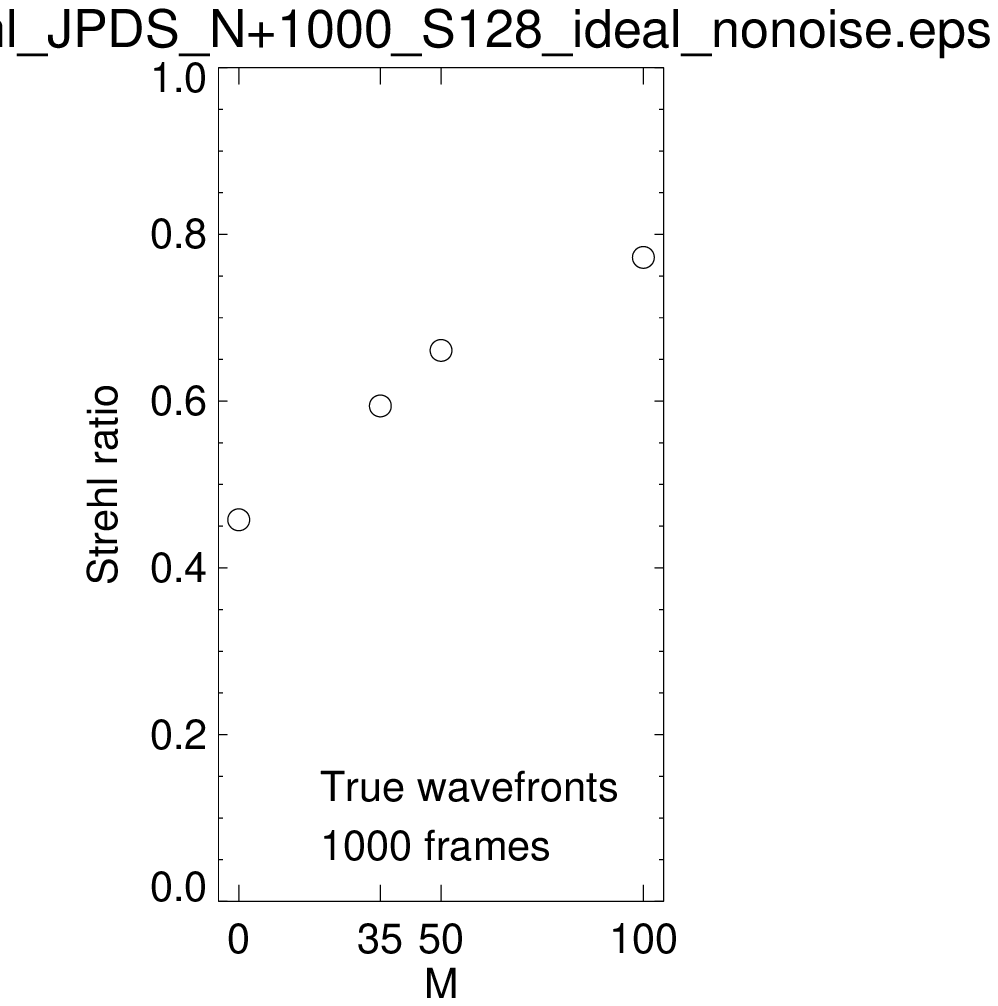}}
  \subfloat[]{\includegraphics[bb=38 24 196 283,clip,width=0.16\linewidth]{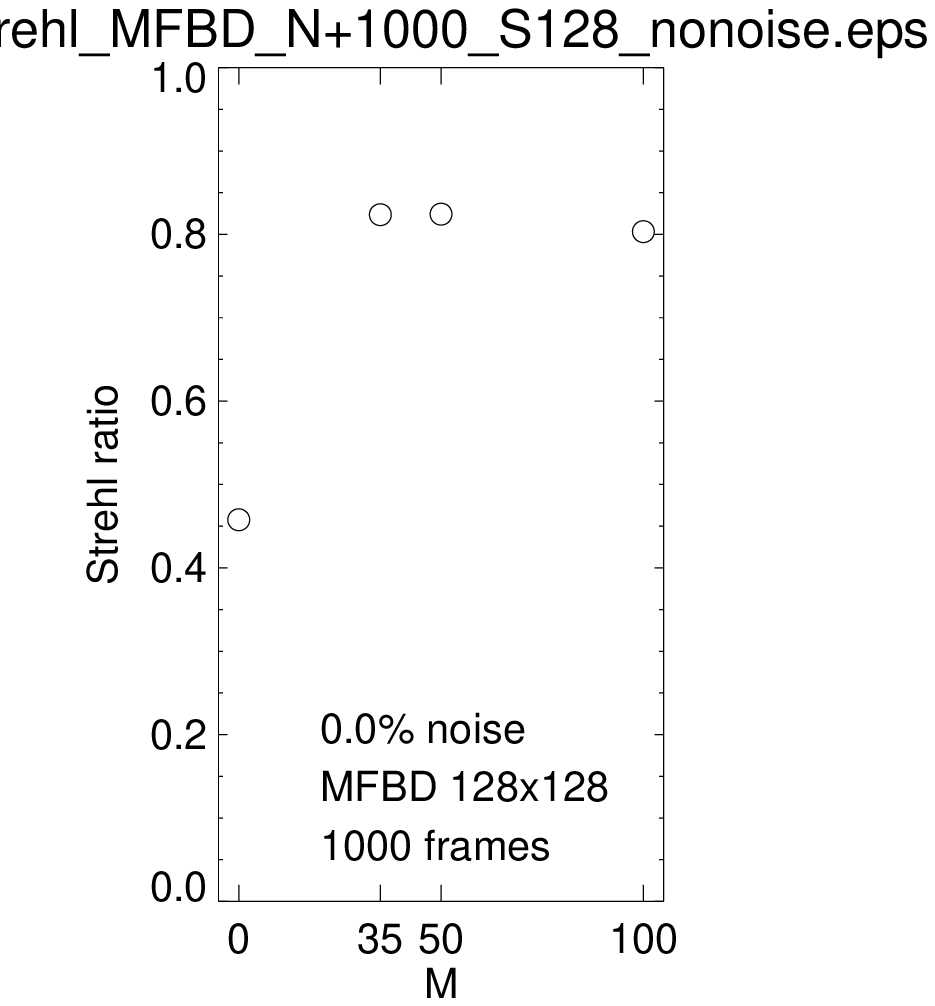}}
  \subfloat[]{\includegraphics[bb=38 24 196 283,clip,width=0.16\linewidth]{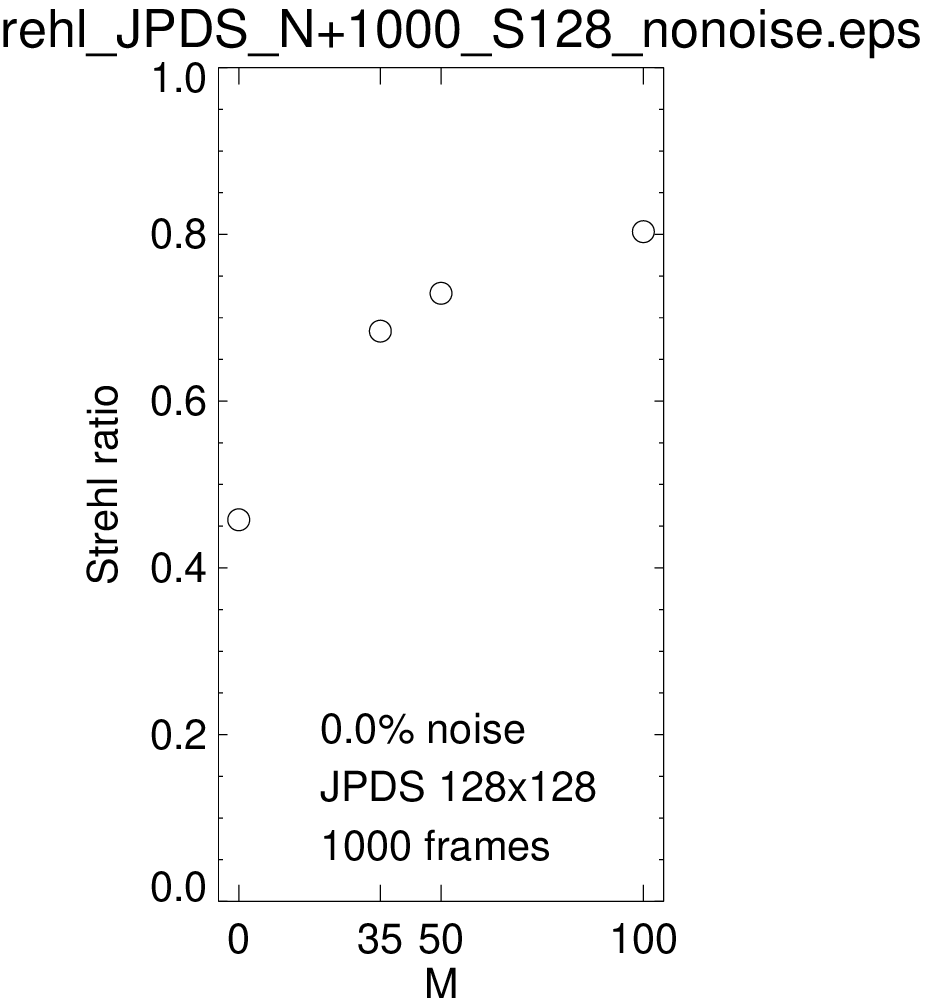}}
  \subfloat[]{\includegraphics[bb=38 24 196 283,clip,width=0.16\linewidth]{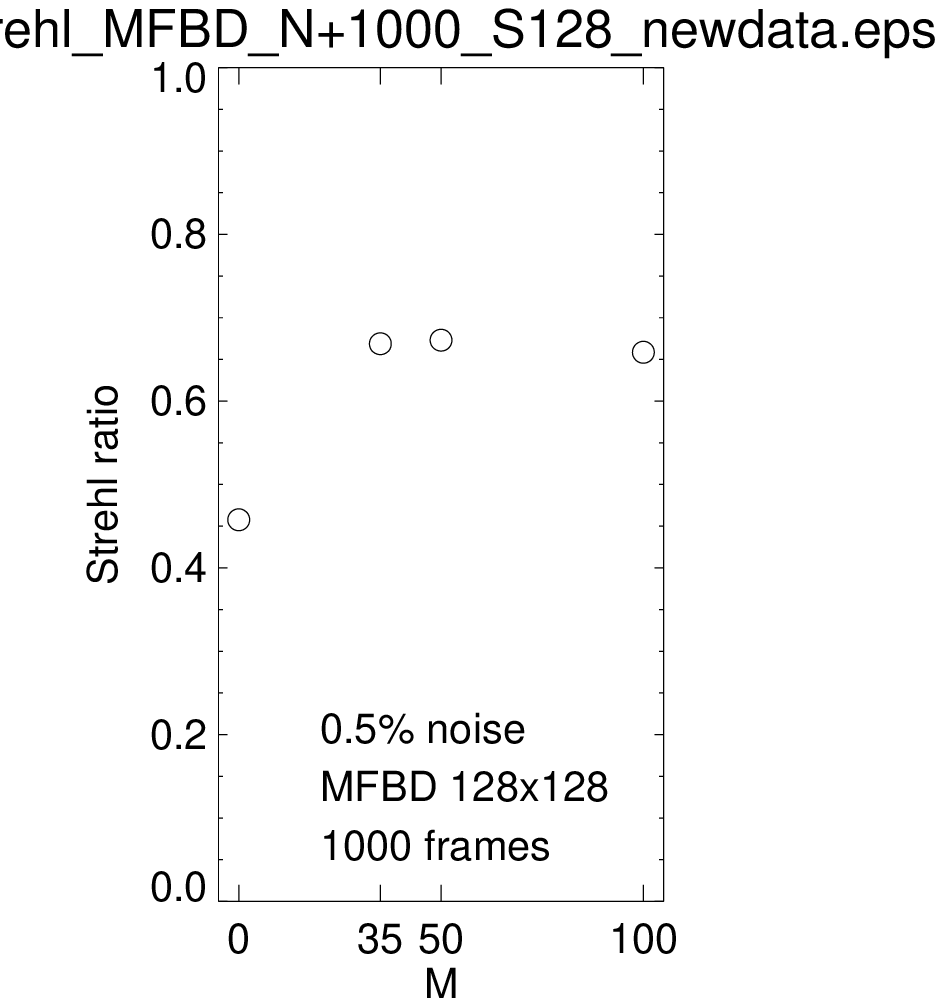}}
  \subfloat[]{\includegraphics[bb=38 24 196 283,clip,width=0.16\linewidth]{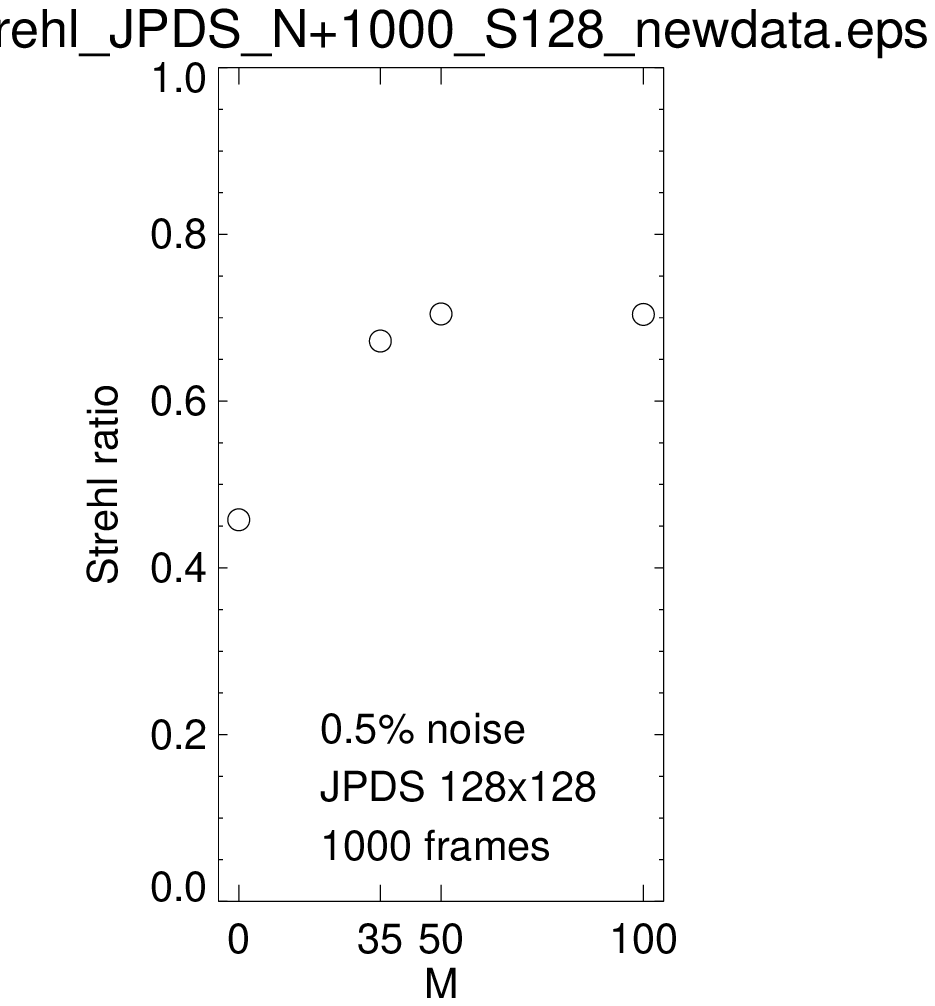}}
  \caption{Strehl ratio as a function of $M$ (number of corrected
    modes) for $r_0=10$~cm.}
  \label{fig:final_strehl}
\end{figure*}

\subsubsection{Results without $\hat S$ compensation}
\label{sec:without-shat}

Figure\@~\ref{fig:final} shows RMS intensity errors of restored images
using different techniques (MFBD, JPDS), number of aberration
parameters ($M=35$, 50 and 100) with and without $\hat S$ compensation
for different values of Fried's parameter $r_0$. We emphasize that for
all calculations made, 1000 frames were used to obtain the wavefronts
but in panels d--f and j--l subsets of only 10 frames were used to
restore the images. This corresponds to processing of SST/CRISP data,
where the broad-band channel of a MFBD or JPDS data set corresponds to
on the order of 500--1000 images. The simultaneously exposed
narrow-band CRISP images are divided into typically 10--12 wavelength
bins, each with 4 polarization bins. These narrow-band images are
restored individually, but using the aberrations (mainly) determined
by the 1000 broadband images. The upper row of plots show the results
obtained without noise and the bottom row show the corresponding
results with 0.5\% Gaussian noise added to the simulated images.

Discussing first the results without $\hat S$ compensation (blue
symbols), our reference for comparison is the results obtained using
1000 frames to restore the image and the first $M$ true (exact)
aberration parameters (panel a). This shows the expected behavior: The
quality of the restored images improves with increasing value of $r_0$
and also with increasing number of estimated aberration parameters
used to restore the images. When $r_0$ is 20~cm, the RMS error with 50
perfectly known aberration parameters is only 1.5\%. However, in more
typical seeing conditions ($r_0=10$~cm or smaller), the corresponding
RMS error is 2.5--8\%, depending on the number of aberration
parameters compensated and the seeing quality.

Figure~\ref{fig:final}b (MFBD processing with 1000 frames)
demonstrates that the RMS intensity error obtained using only focused
images to estimate the aberration parameters shows quite small
\emph{variation} of the intensity error with the number of aberration
parameters $M$, when $M$ is in the range 35--100. The RMS intensity
error for $M=35$ MFBD calculations actually corresponds to what was
obtained with the \emph{true} aberration parameters for $M=100$ (Panel
a)! \emph{This implies that MFBD processing with a given number of
  aberration parameters leads to estimates of the transfer function
  that compensate for the effects of missing higher-order aberration
  parameters (truncation of the wavefront).} Increasing the number of estimated aberration
parameters from 35 to 50 or 100, reduces the efficiency of this
compensation such that the reduction of the intensity error is
relatively modest. Quite clearly, the estimated wavefront with MFBD
processing must be inaccurate, but such that the estimated {\em
  transfer function} accurately represents the true transfer function.
This conclusion is further supported by Fig.\@~\ref{fig:final_phi},
which shows the variation of the (estimated) wavefront RMS and
wavefront error for different number of estimated wavefront parameters
when $r_0=10$~cm. Panel a (red symbols) shows that the RMS of the
\emph{true} wavefront increases by 50\% when $M$ is increased from 35
to 100. In contrast, the RMS of the \emph{estimated} wavefront for the
$M$ first aberration parameters is strongly overestimated for $M=35$
and $M=50$ with MFBD processing such that the estimated wavefront RMS
is nearly independent of $M$. Thus, using a low-order or high-order
representation for the wavefront leads to nearly the same estimated
wavefront RMS. A striking result is that MFBD processing of the first
35--100 corrected modes leads to residual wavefront errors (blue
symbols) that are larger than if the wavefront is estimated to be
exactly zero!

When using also a defocused channel (JPDS processing), the wavefronts
are more constrained to represent reality, leading to smaller
wavefront errors (Fig.\@~\ref{fig:final_phi}c). But this constraint
also limits the freedom in compensating the transfer function for
high-order aberrations, such that \emph{larger} intensity errors are
obtained in the restored images with JPDS (Fig.\@~\ref{fig:final}c)
than with MFBD processing (Fig.\@~\ref{fig:final}b).

These conclusions about differences between MFBD and JPDS processing
are supported by the Strehl ratios calculated from wavefronts
determined with the granulation images and shown in
Fig.\@~\ref{fig:final_strehl}. With true aberration parameters
(Fig.\@~\ref{fig:final_strehl}a) and without $\hat S$ compensation,
the Strehl intensity increases gradually with the number of
compensated aberration parameters $M$. However, with MFBD processing,
the Strehl ratio is constant when $M$ is in the range 35--100. The
Strehl ratio achieved, $R=0.8$, is close to what is expected from
Eqs.\@~(\ref{eq:9}) and (\ref{eq:10}) for $N=M=100$. The variation of
the Strehl ratio with $M$ obtained with JPDS processing is
intermediate to that of MFBD and using the true aberration parameters.

Adding 0.5\% Gaussian noise to the MFBD data shows that the results are
sensitive to noise (Figs.\@~\ref{fig:final}h--i) and indicates that in
practice the compensation effects discussed above will not be large
and that the quality of MFBD and JPDS image restorations should be
rather similar. The RMS intensity error improvement is limited to that
corresponding to about 50 ``true'' aberration parameters, or a Strehl
ratio of 0.7. Using more than 50 aberration parameters to represent
the wavefront does not significantly improve the quality of the
restored images.

\subsubsection{Results with $\hat S$ compensation}
\begin{figure*}[!t]
  \centering
  \subfloat[]{\includegraphics[angle=0,width=0.24\linewidth]{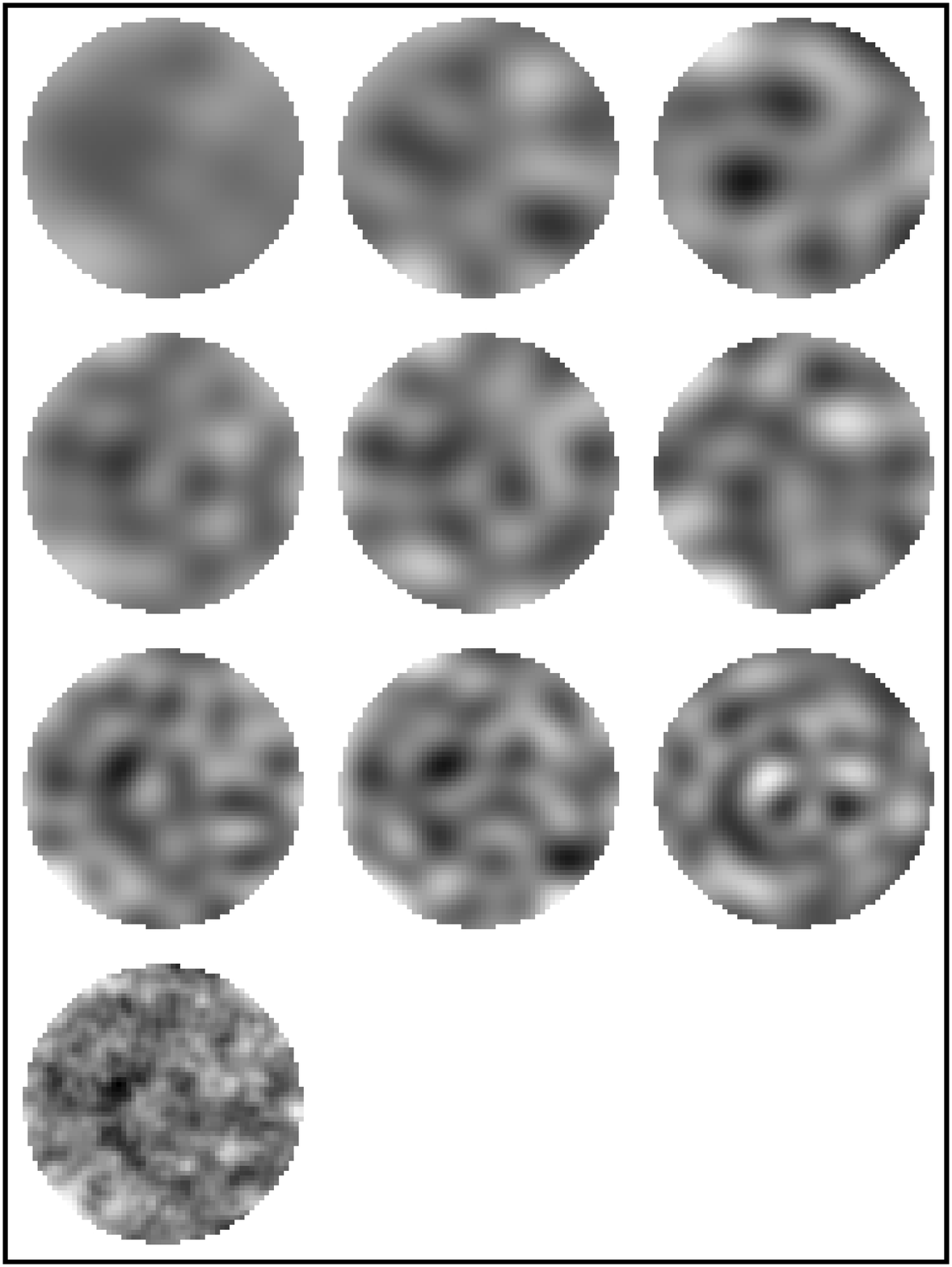}}
  \subfloat[]{\includegraphics[angle=0,width=0.24\linewidth]{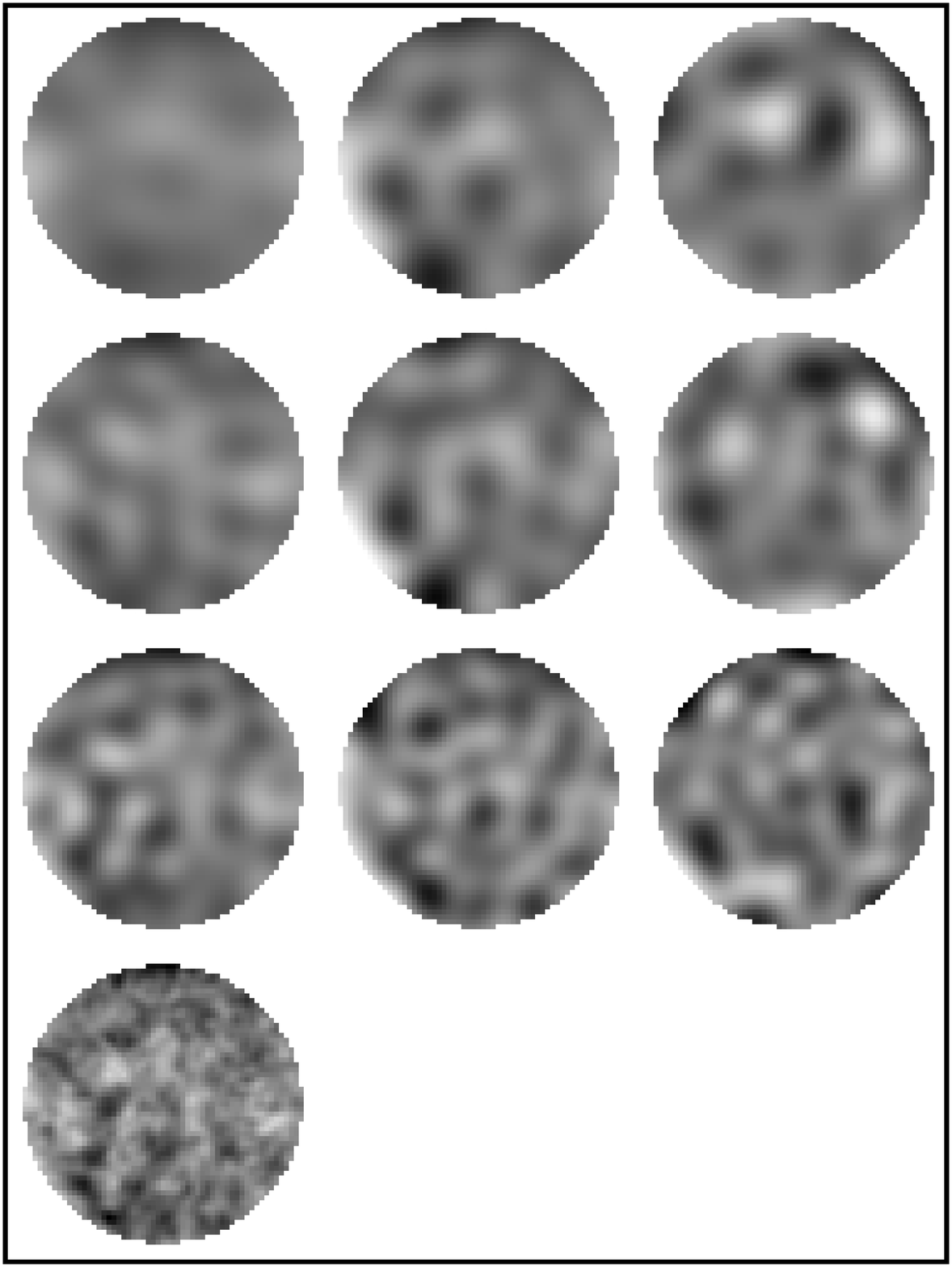}}
  \subfloat[]{\includegraphics[angle=0,width=0.24\linewidth]{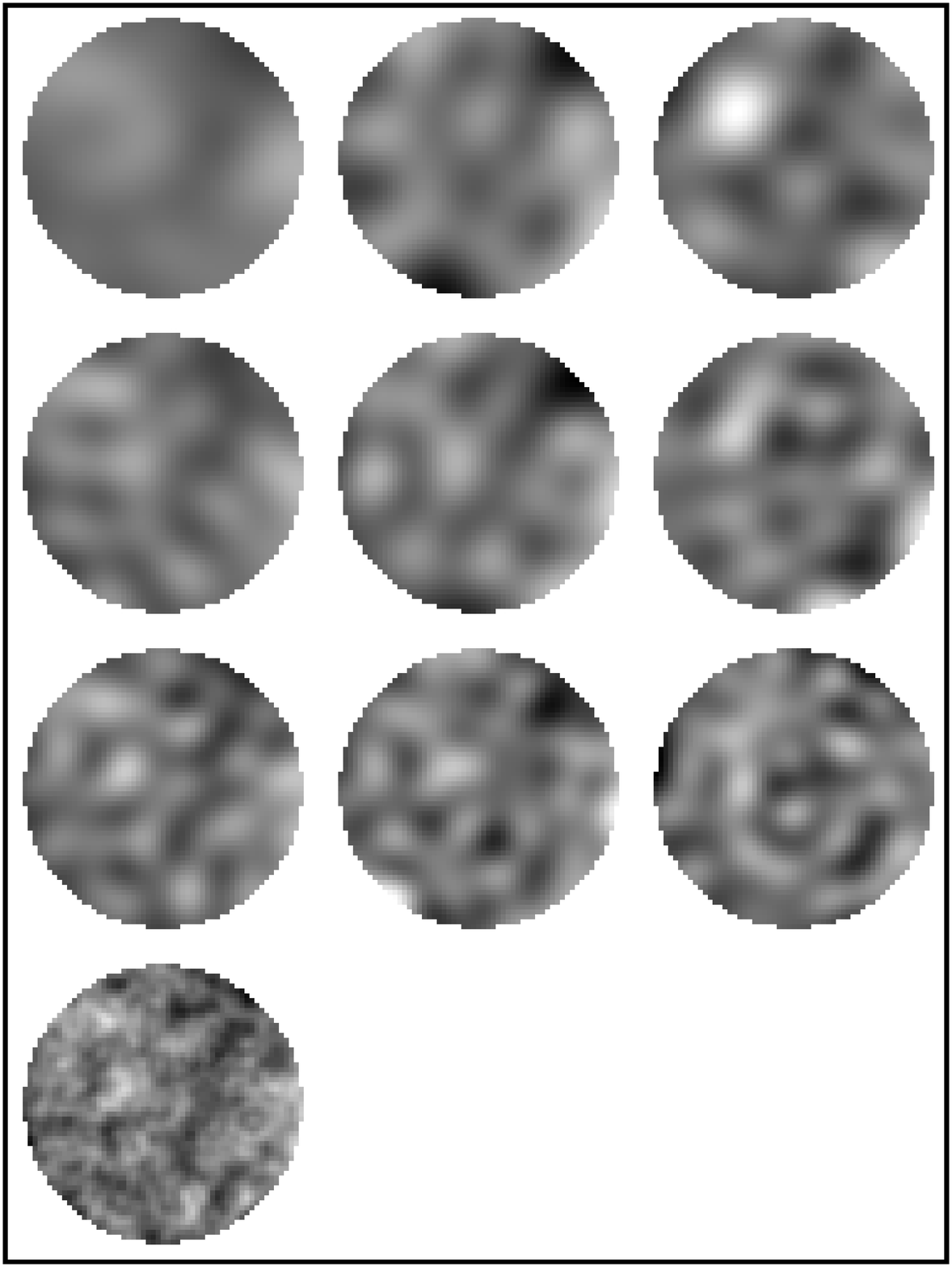}}
  \subfloat[]{\includegraphics[angle=0,width=0.24\linewidth]{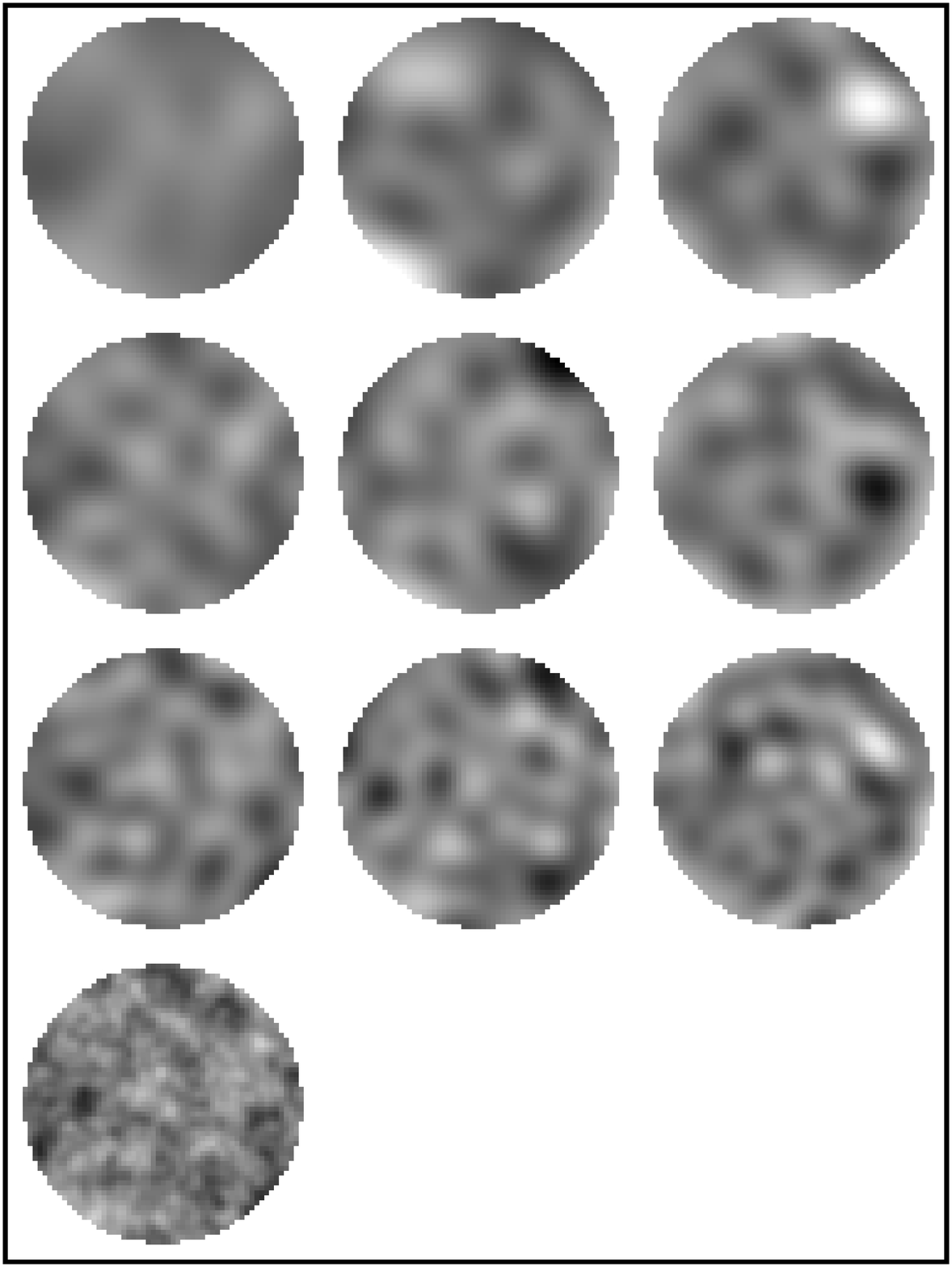}}
  \caption{Sample wavefronts from the simulation experiments.
    \textbf{(a)}--\textbf{(d)}:~Different realizations of the
    Kolmogorov statistics. Within each sub-figure, all wavefronts are
    displayed with identical scaling. \textbf{Left}:~True wavefronts;
    \textbf{Center}:~JPDS estimated wavefronts; \textbf{Right}:~MFBD
    estimated wavefronts. \textbf{1st row}:~35 KL modes; \textbf{2nd
      row}:~50 modes; \textbf{3rd row}:~100 modes; \textbf{4th
      row}:~All 1001 modes.}
  \label{fig:phases}
\end{figure*}
The results obtained with $\hat S$ compensation are shown with red
symbols in Fig.\@~\ref{fig:final}. With such compensation,
\emph{restorations with ``true'' low-order wavefronts allow nearly
  perfectly restored images, even with 10 frames, in bad seeing
  ($r_0=7$~cm) and using only 35 aberration parameters}. This
re-enforces our earlier conclusion, that details of the high-order
part of the wavefront are not important as long as it has the
``right'' statistical properties. In particular, the ``speckles'' seen
in the wings of the PSF of Fig.\@~\ref{fig:PSFs} (upper row), have a
small effect on the restored images. The excellent result obtained
with $\hat S$ compensation based on ``true'' wavefronts also is
consistent with the results discussed previously: that MFBD processing
with 35 aberration parameters can produce transfer functions
equivalent to those obtained with compensation by 100 aberration
parameters, even though the actual wavefronts derived are
demonstratedly wrong.
 
With MFBD processing, the improvement from $\hat S$ compensation
varies strongly with the number of aberration parameters, $M$. The
remarkably good results obtained without $\hat S$ compensation for
$M=35$ and $M=50$ constitutes a \emph{problem} with $\hat S$
compensation! This is because our calculations of $\hat S$ do not take
into account the compensation for high-order aberrations already
provided by the MFBD (and JPDS) processing. This leads to an
\emph{overcompensation of the effects of high-order aberrations} for
$M=35$ and $M=50$. This compensation is much smaller with $M=100$.
With JPDS processing (panels c, f, i and l), the results show improved
consistency. Here, $\hat S$ compensation with $M=50$ or $M=100$ leads
to RMS intensity errors that are reduced by a factor 2--3. The
obtained RMS errors are consistently smaller than for MFBD processing
without $\hat S$ compensation, when noise is present.
 
We have here demonstrated that MFBD image reconstruction by itself
allows compensation effects by fitting transfer functions rather than
wavefronts. In this sense, MFBD or JPDS processing provides more
optimum image reconstruction than with a Shack-Hartmann based
wavefront sensor. The main problem in the present context is that MFBD
and (to a smaller extent) JPDS processing already introduces part of
the compensation intended to be performed with $\hat S$. This leads to
overcompensation if no counter measures are taken. By using JPDS
processing, the compensation effects are constrained and the risk of
overcompensation with $\hat S$ is small. 

The relatively poor estimates of the \emph{wavefronts} obtained with
MFBD and JPDS processing may appear to contradict earlier simulations
with phase diversity methods
\citep{lofdahl94wavefront,paxman96evaluation}, but are a direct
consequence of the assumed AO correction, reducing the RMS amplitudes
of the first 35 KL aberrations with a factor of 4. This leaves
residual wavefronts that have small or negligible amplitudes for the
first 35 modes and a correspondingly large contribution from
higher-order modes. This leads to significant cross-talk from
high-order aberrations, degrading the low-order wavefront estimates
but actually leading to better estimates of the transfer functions
than expected on the basis of the number of aberration parameters
included. Figure~\ref{fig:phases} shows the wavefronts corresponding
to the first 4 frames used to reconstruct the image with 10 frames. In
the left column are shown (top to bottom) the true wavefronts
represented with 35, 50, 100 and 1001 KL modes. It is evident from the
lower-left wavefront in each panel that the true wavefront has been
stripped of its low-frequency wavefront components. It is also evident
that the JPDS (mid column) and MFBD (right column) wavefront estimates
show large differences and that JPDS provides the best estimate of the
true wavefront.

\subsection{Inversion tests with point sources}
\label{sec:invers-test2}

We have also made inversion tests with point sources. While not
directly relevant for restoration of solar images, these tests do shed
further light on the MFBD and JPDS compensation effects discussed
above. The most striking result is that both MFBD and JPDS processing
of synthetic noise-free point source images lead to nearly perfectly
reproduced low-order wavefronts.
To understand why granulation images and images of point sources lead
to different wavefront estimates, we express the equation used to
estimate the transfer function as follows: Inserting the expression
for $\hat F$ in Eq.\@~(\ref{eq:2}) into Eq.\@~(\ref{eq:1b}), and
replacing $D_k$ with $F T_k$, we obtain an expression for the error
metric $L$ that corresponds to the scalar quantity minimized in MFBD
and JPDS processing:
\begin{equation}
  L = \sum\limits_{u,v} |F|^2 \sum\limits_{k=1}^{K} \biggl|T_k - 
  {\hat T}_k \sum\limits_{n=1}^{K} T_n {\hat
    T}_n^*\biggm/\sum\limits_{n=1}^{K}{|{\hat T}_n|^2} \biggr|^2 
\end{equation}
Ignoring the complicated expression involving the exact and estimated
transfer functions $T_k$ and ${\hat T}_k$, we emphasize that this
expression is multiplied by the power spectrum of the object, $|F|^2$.
For a point source, this power spectrum is unity at all spatial
frequencies but for solar fine structure, the power spectrum falls off
at high spatial frequencies. Depending on the object re-imaged,
smaller or larger weight will be given to the high spatial
frequencies, obviously leading to different results as regards the
derived wavefronts but actually quite small differences in the derived
transfer functions.

\subsection{Comment on the use of small subfields}

The results discussed above were all obtained with
128$\times$128-pixel subfields. Calculations with 256$\times$256-pixel
subfields gave results that are virtually identical to those
discussed. MFBD calculations with 80$\times$80-pixel subfields gave
poorer, but still acceptable, results. However, JPDS restorations with
80$\times$80-pixel subfields essentially failed. This is attributed to
the relatively large diameter of the PSF corresponding to the
\emph{defocused} images, causing problems from lack of information
about the object outside the subfield and wrap-around effects when
using FFTs to perform convolutions with the defocused PSF. It is thus
more difficult to use small subfields with JPDS than with MFBD with
the present methods.

\section{Tests with real data}
\label{sec:tests-with-real}

\subsection{Observations}

To test the proposed $\hat S$ compensation method we used granulation
images recorded through a 0.44~nm FWHM 2-cavity filter centered at
630.26~nm and a 0.34~nm filter centered at 538.20~nm, both used as
wide-band channels for the SST/CRISP imaging spectro-polarimeter
\citep{scharmer08crisp}. The images were exposed by means of a
rotating chopper, set to give exposure times of 16~ms and a dark
read-out time of nearly 12 ms, corresponding to an overall frame rate
of 36~Hz. The 630~nm and 538~nm data discussed here were recorded between 10:45 and 10:50 UT on 26 June 2009, during a period with reasonably good but strongly variable seeing. The science target for these observations was a small
pore in AR 1023, located at approximately S22, W20, corresponding to a
heliocentric distance of about $\theta=30\degr$
($\mu=\cos\theta=0.87$). This pore was also used as lock point for the AO system. However, in the present paper we discuss only field-free granulation outside this active region. The noise level in the recorded images was estimated from power spectra at high spatial frequencies and found to
be about 0.9\%.

CRISP was used to repeatedly scan each of the 630.26~nm and 538.20~nm
lines three times, using different numbers of wavelength positions for
the two lines. These scans required a total of 17~s for the 630~nm line and 12~s for the 538~nm line, setting the time between pre-filter changes. Simultaneously, approximately 600 wide-band
images were collected at 630~nm and $\sim$433 images at 538~nm during
each scan. We refer to these 17~s and 12~s sets of wide-band images as ``full scan'' data sets but will primarily discuss the results of processing 2.2~s sub-sets of this data to match the time-scale of seeing changes.

The images were corrected for gain and bias and MFBD processed with
the MOMFBD code, see Sect.\@~\ref{sec:data-processing} below.

\subsection{Seeing measurements}
\begin{figure}[!t]
  \centering
  \includegraphics[bb=10 10 493 343,clip,width=\linewidth]{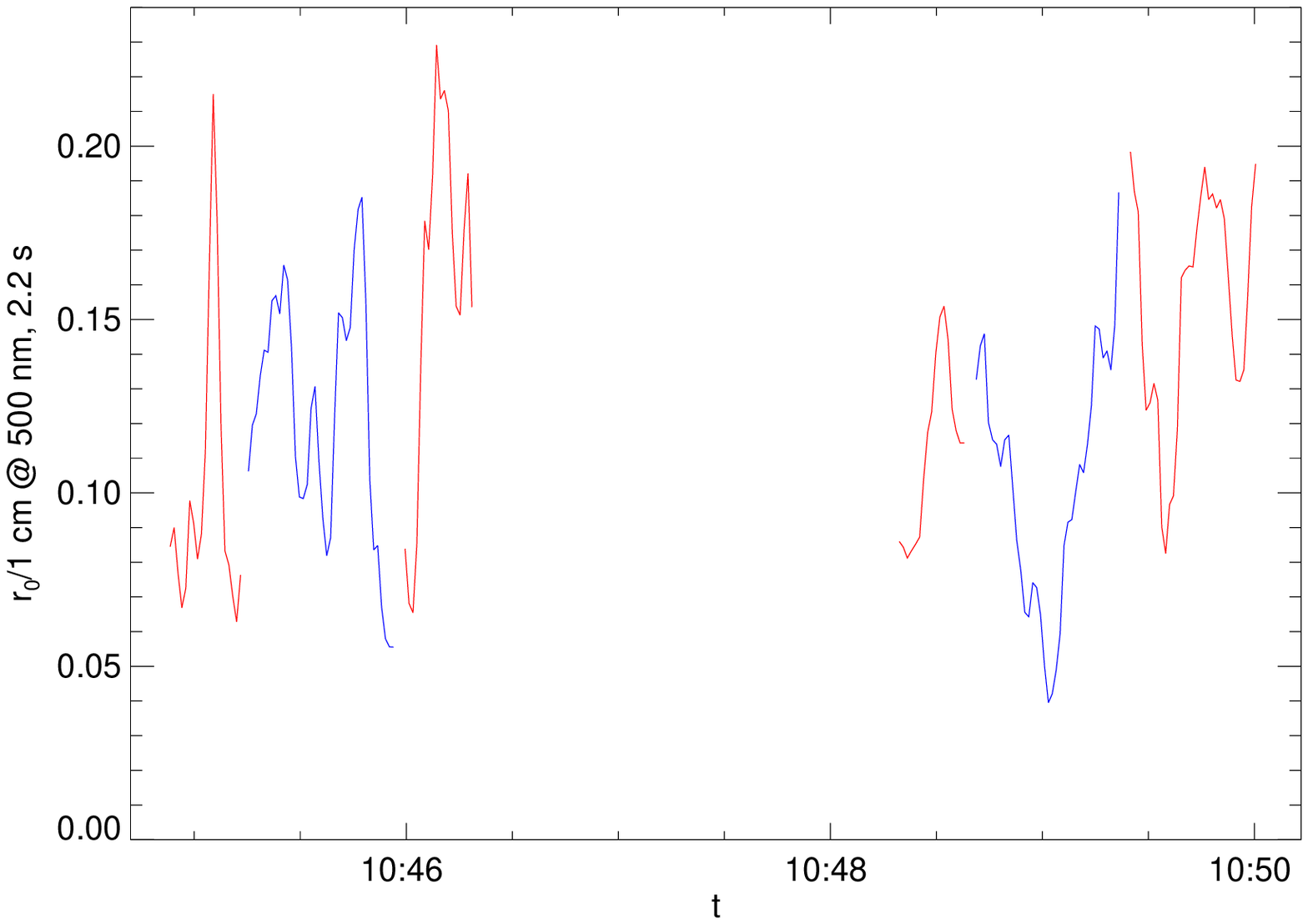}
  \includegraphics[bb=10 10 493 343,clip,width=\linewidth]{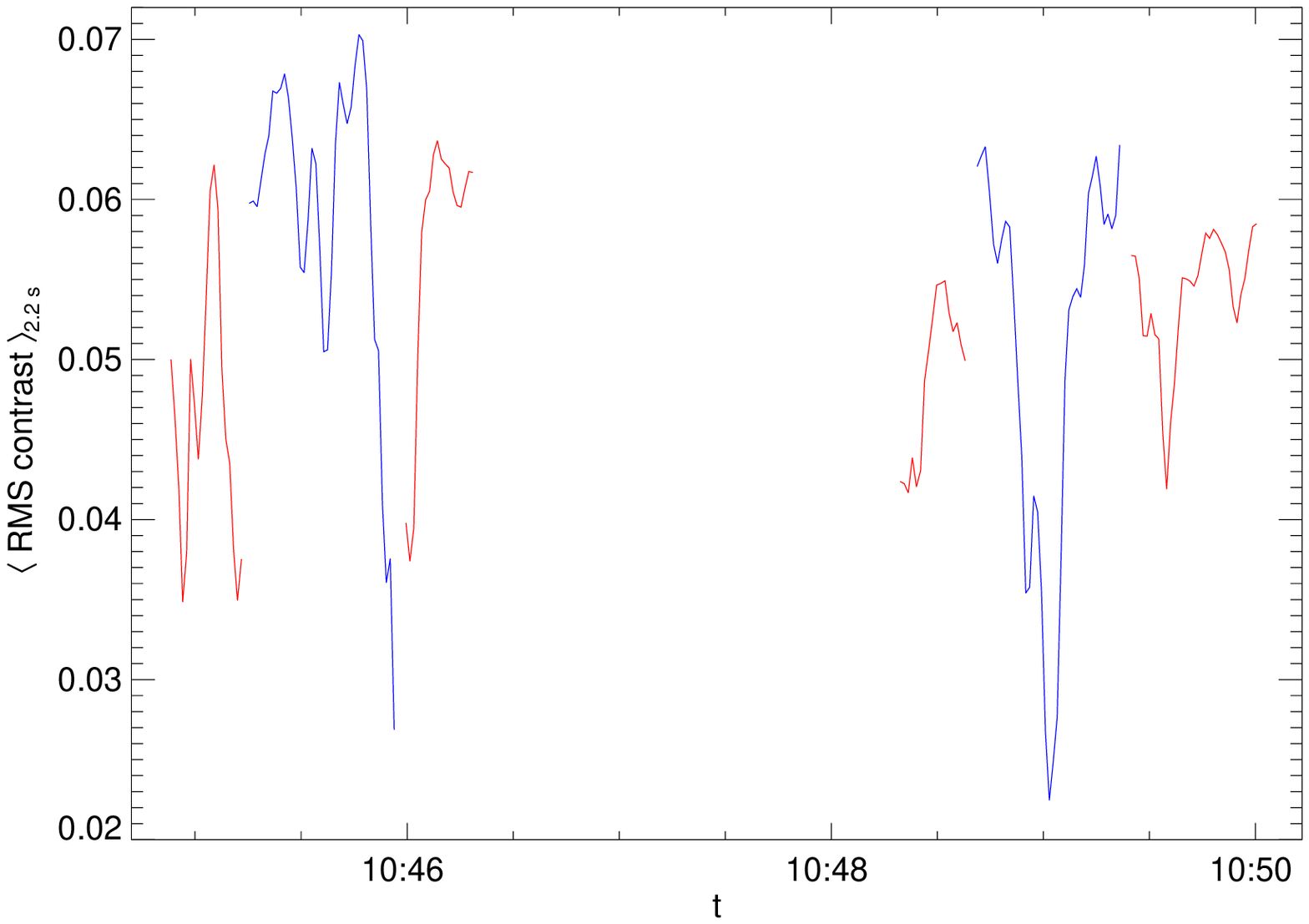}
  \includegraphics[bb=10 9 493 343,clip,width=\linewidth]{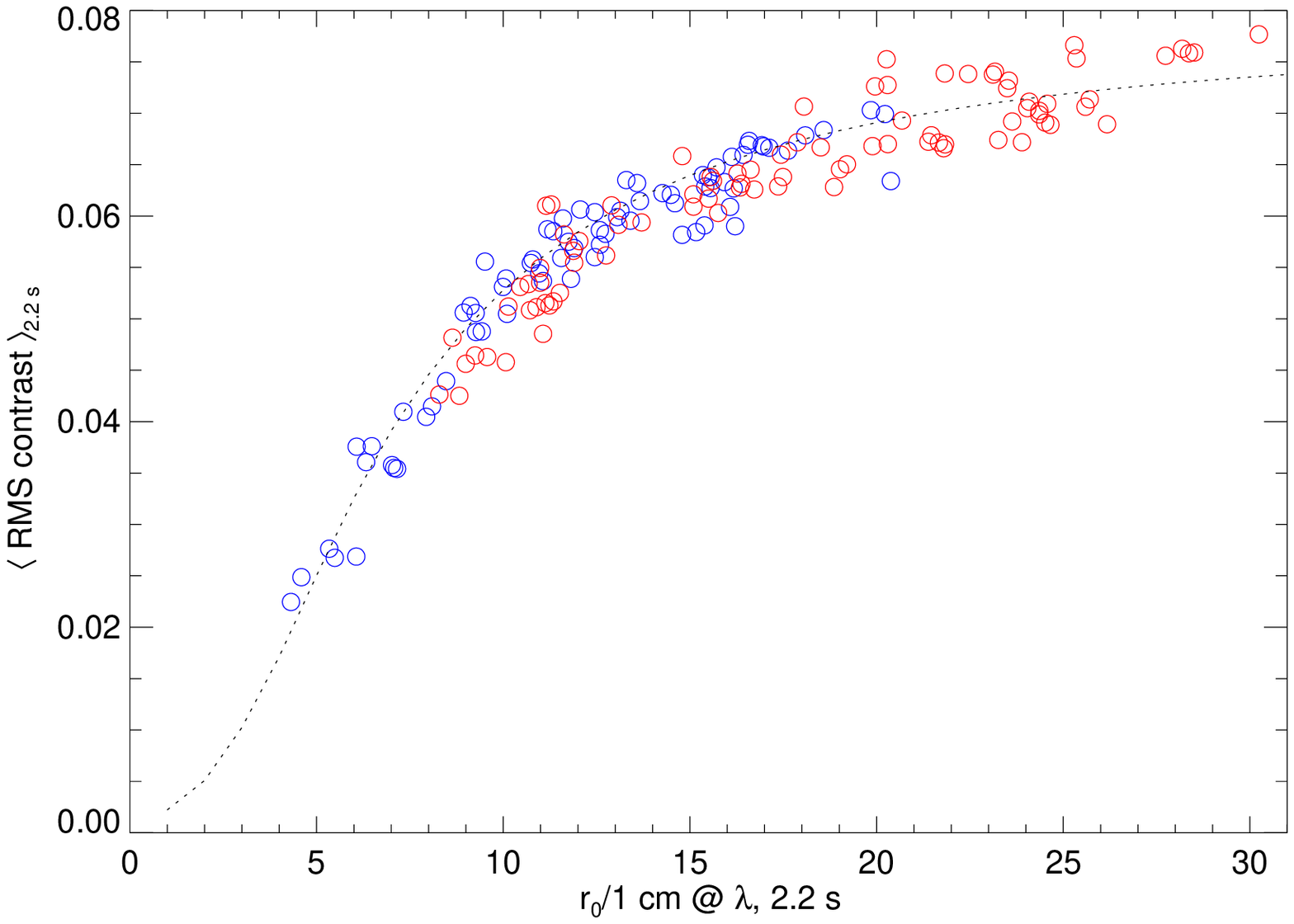}
  \caption{Variation of Fried's parameter $r_0$ with time, measured
    with the SST wide-field wavefront sensor bypassing the AO system
    (top panel) and the corresponding variation of the measured
    granulation RMS contrast of the science images (mid panel)
    obtained at $\lambda=538$~nm (blue curves) and at $\lambda=630$~nm
    (red curves) with the AO system in closed loop. The bottom panel
    shows the correlation between $r_0$ and the RMS contrast at the
    two wavelengths but with the contrast at $\lambda=630$~nm
    multiplied by 1.22 to bring the data onto a common curve. The
    dotted curve corresponds to the contrast of simulated data shown
    in Fig.\@~5, divided by a factor 1.85.}
  \label{fig:scatter_r0_raw}
\end{figure}
Simultaneous measurements of Fried's parameter $r_0$ were obtained
with the SST wide-field wavefront sensor \citep[WFWFS;
][]{2010A&A...513A..25S}, developed as part of an effort to characterize
day-time high-altitude seeing at La Palma. The WFWFS is mounted on a
beam that splits off light immediately before the tip-tilt and
deformable mirrors, such that $r_0$ can be measured through the
telescope and from a FOV that is adjacent to the science FOV, but
without impact from the AO system. Between UT 10:00 and 10:48,
processed WFWFS data averaged over 100~s gave estimates of $r_0$ in
the range 11--12~cm at 500~nm, corresponding to 15--16~cm at 630~nm
and 12--13~cm at 538~nm. However, the science images, recorded with
the AO system in closed loop, showed strong variations in image
quality on time scales on the order of a few seconds. The WFWFS variances measured are proportional to $r_0^{-5/3}$ \citep[Eqs. (7)--(8);][]{2010A&A...513A..25S}, such that averaging wavefront sensor data in variable seeing gives the largest weights to relatively poor seeing (small
values of $r_0$). However, MFBD image restoration gives the highest
weights to good images (large values of $r_0$). We therefore
compare data recorded over such short time intervals that
$r_0$ can be considered constant. We re-processed the WFWFS data in
blocks of 2.2~sec, or about 10 WFWFS CCD frames. Figure
\@~\ref{fig:scatter_r0_raw} shows in the upper panel the variation of
$r_0$ with time from UT 10:45-10:50, obtained from overlapping 2.2-sec
blocks of WFWFS data. The red and blue curves in this panel indicate
the wavelengths at which the science images were recorded, at 630~nm
and 538~nm respectively. In the mid panel is shown the variation of
the RMS contrast measured (see Sect.\@~\ref{sec:granulation-contrast}) for the 630~nm and 538~nm images
after flat-fielding but without MOMFBD image restoration. Comparing
the two panels, there is a clear correlation between $r_0$ and the
measured RMS contrast; even small and rapid variations are reproduced
in detail. The bottom panel shows the correlation between all measured
values of $r_0$ and the RMS contrast. In this figure, we have scaled
the $r_0$ values measured with the WFWFS at 500~nm to 538~nm and
630~nm by assuming that $r_0$ is proportional to $\lambda^{6/5}$. We
have also multiplied all RMS contrasts at 630~nm by an ad hoc factor
1.22, roughly compensating for the wavelength dependence of the
granulation contrast. The plotted values cover a range of $r_0$ from
4~cm to nearly 30~cm and RMS contrast from 2.2\% to 7.8\% and show an
excellent correlation. The dotted curve corresponds to variation of
the RMS contrast with $r_0$ obtained from the simulations shown in
Fig.\@~\ref{fig:images}, but divided by a factor 1.85 to fit the data. This large
factor in part comes from the assumed perfect compensation of the
first 35 KL modes, corresponding to an AO system with 100\%
efficiency, and in part must be due to stray light.

Our conclusion from the excellent correlation between measured
variations of $r_0$ with the WFWFS (bypassing the AO system) and the
science images (recorded with the AO system in closed loop) is that
the WFWFS indeed provides an accurate measure of seeing, although we
cannot from this data rule out systematic errors, such as a scale
factor error or a systematic bias, in the measured $r_0$ values.
 
\subsection{MFBD processing}
\label{sec:data-processing}

The MOMFBD software package was used to divide the images into
128$\times$128-pixel overlapping subfields, which were separately MFBD
restored using the $M=100$ most significant KL modes. The restored
subfields were then mosaicked to produce restored images over the
entire observed FOV. To investigate the effects of strongly variable
seeing on the MFBD processing, we processed the data in blocks
corresponding to about 2.2~sec of data as well as in blocks
corresponding to full scans (12~sec resp. 17~sec data). To allow a
comparison between the two methods of processing, subsets of the full
scan MFBD wavefronts and observed images were used to restore the same raw images as used with the 2.2~sec data.

The observed images are over-sampled by 12\% at 630~nm but 4\%
under-sampled at 538~nm. In the MOMFBD code as well as in the $\hat S$
correction, under-sampling is implemented such that the phase over the
entire pupil can be represented but the transfer function $\hat T_k$
is cropped at the Nyquist frequency.

\subsection{Results}

\subsubsection{Granulation contrast measurements}
\label{sec:granulation-contrast}

\begin{figure}[!t]
  \centering
  \subfloat[538~nm]{\includegraphics[angle=90,width=0.48\linewidth]{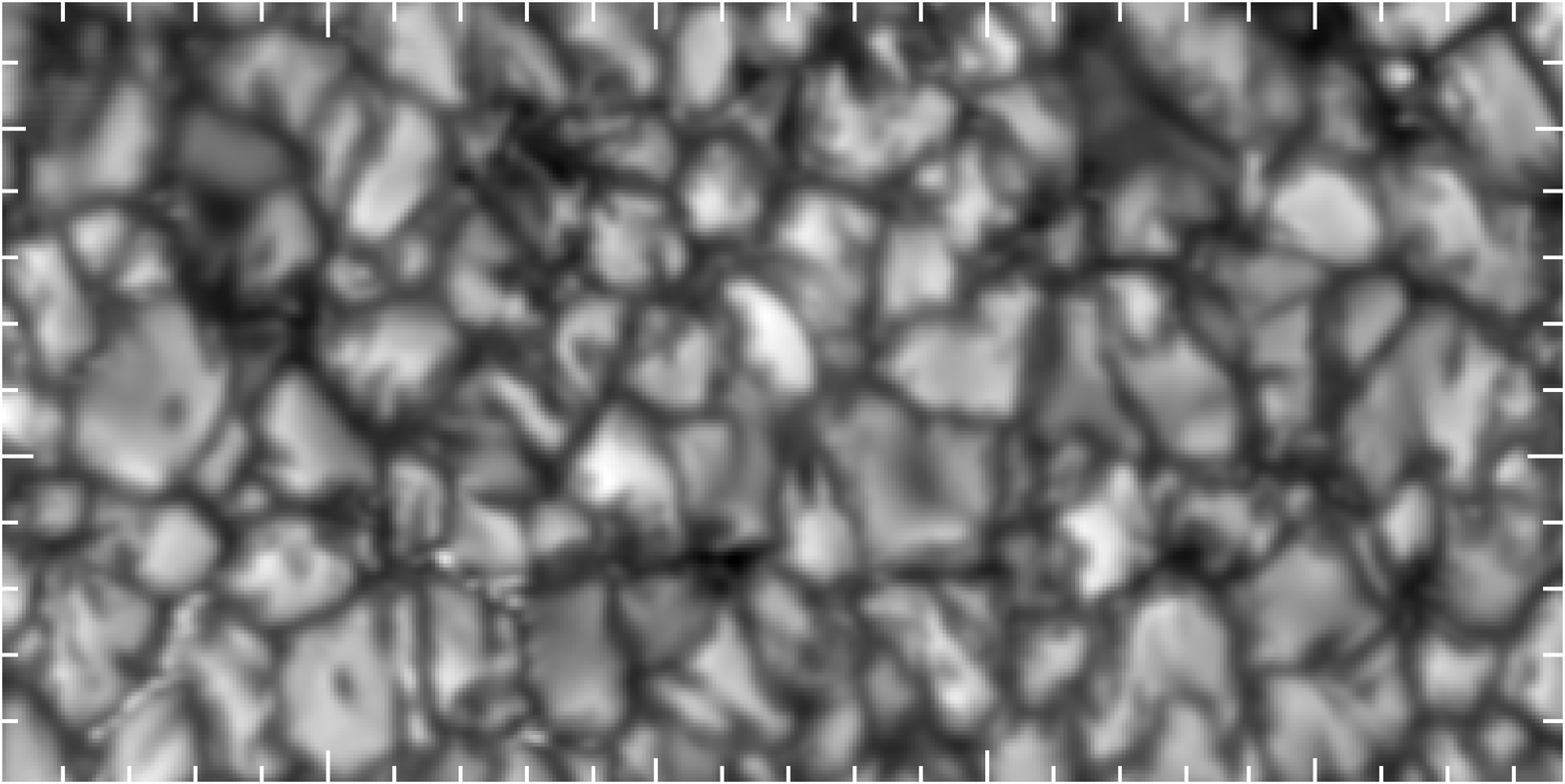}}
  \hfill
  \subfloat[630~nm]{\includegraphics[angle=90,width=0.48\linewidth]{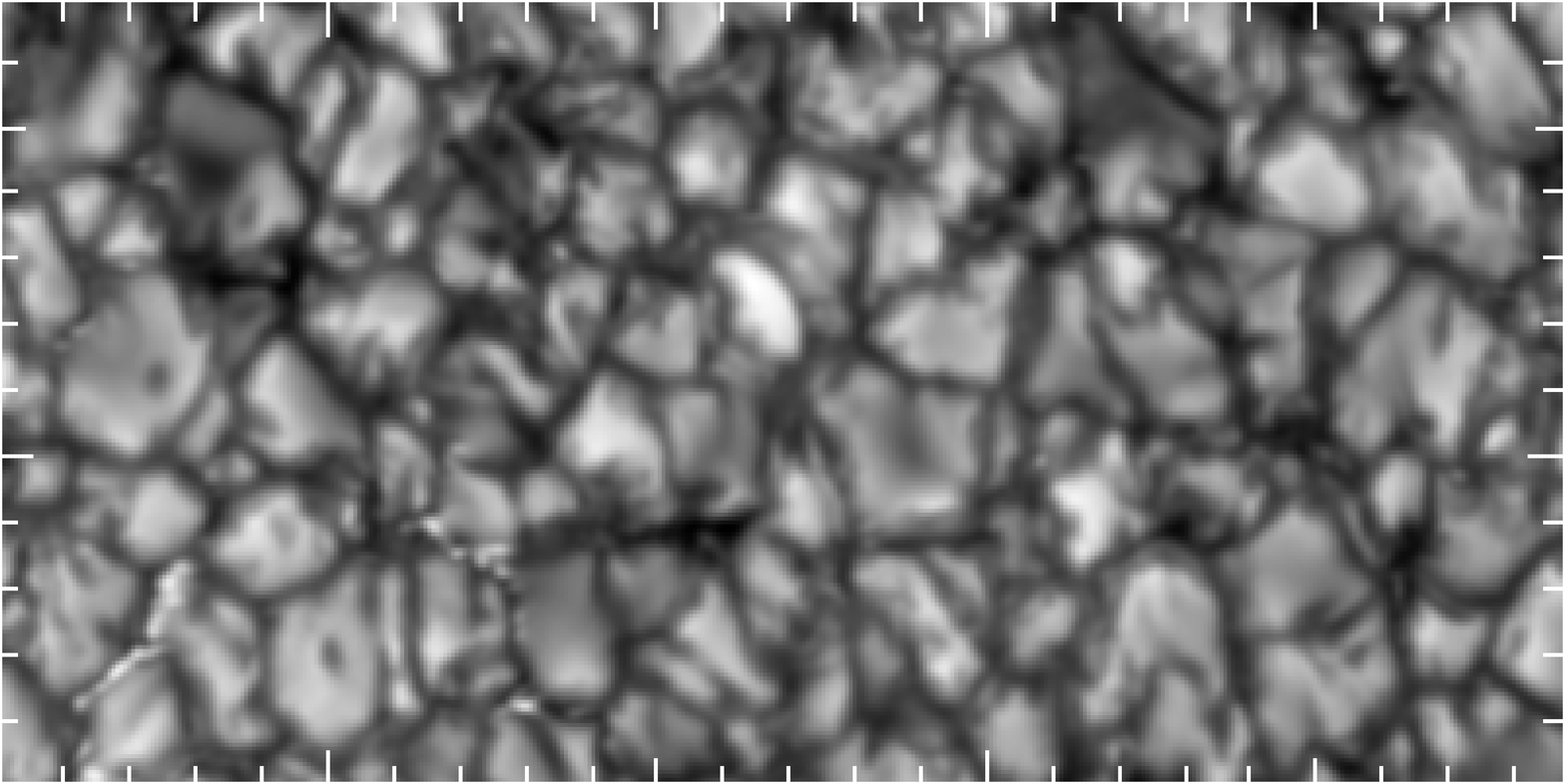}}
  \caption{Examples of MFBD restored images based on data collected at
    $\mu=0.87$ UT 10:50). The 200$\times$400-pixel subfield shown
    was used for RMS contrast measurements. Tick marks are 1\arcsec{}
    apart.} 
  \label{fig:restored}
\end{figure}

\begin{figure}[!t]
  \centering
  \includegraphics[bb=10 9 493 343,clip,width=\linewidth]{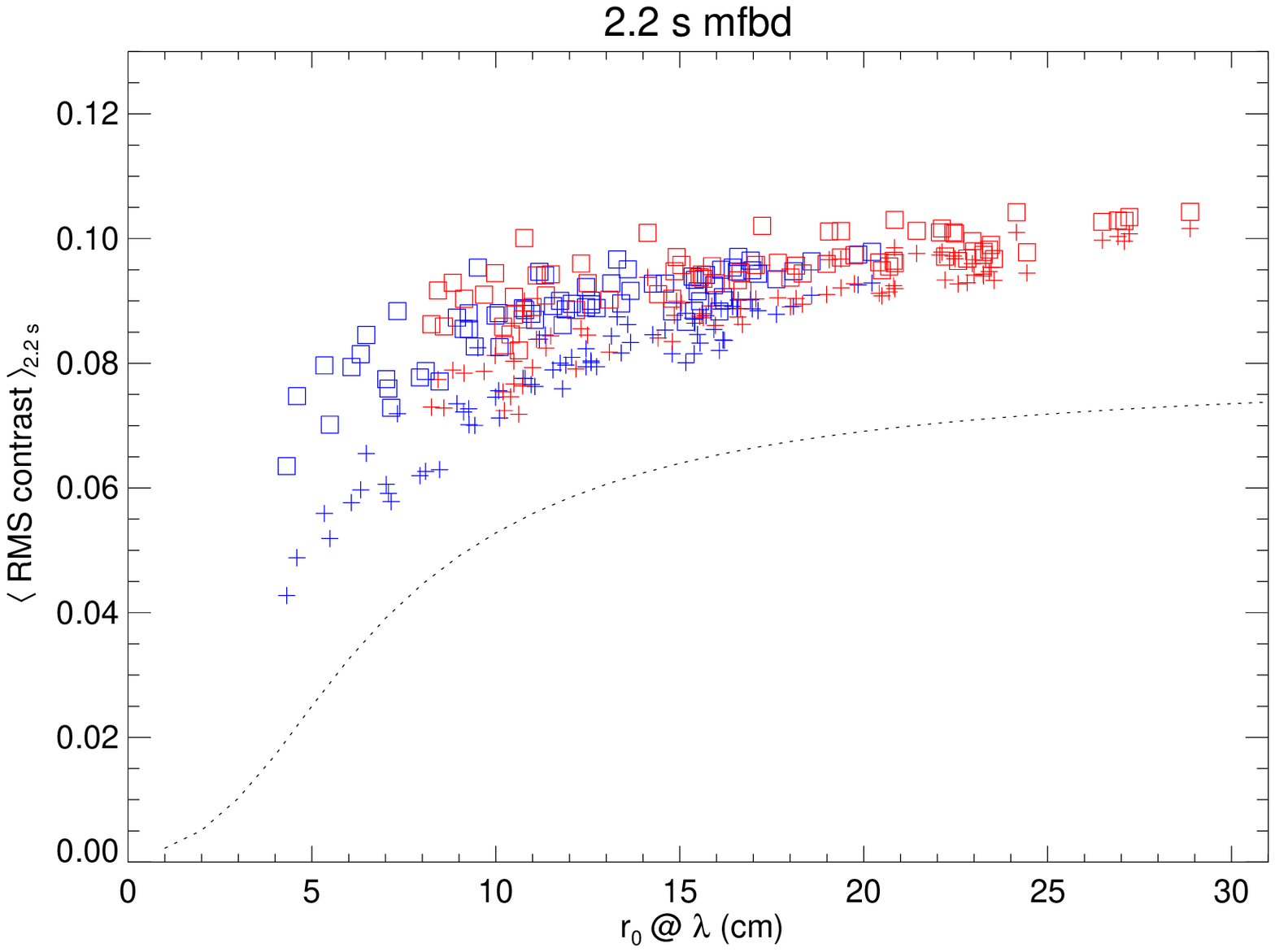}\\[1mm]
  \includegraphics[bb=10 9 493 343,clip,width=\linewidth]{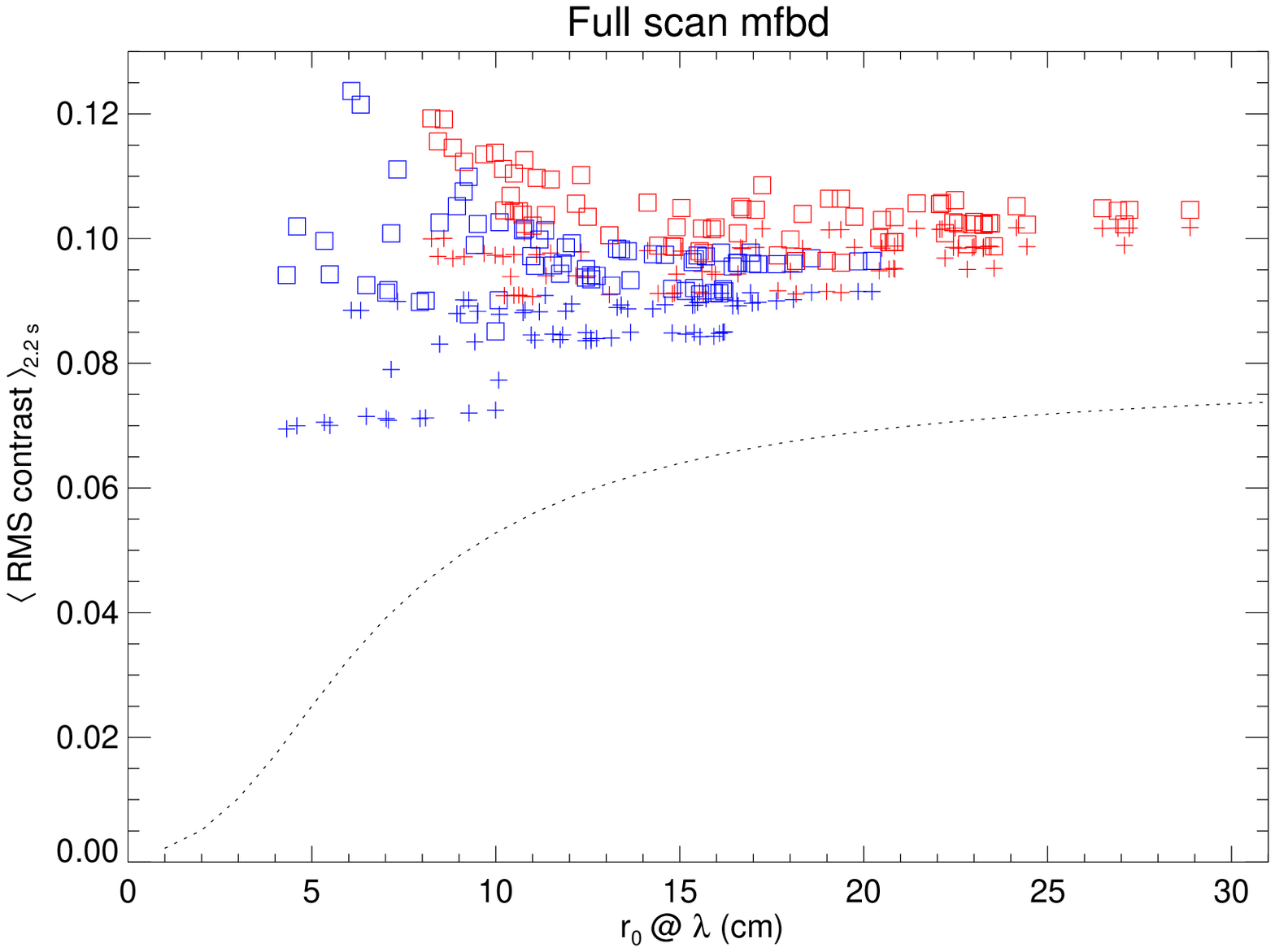}\\[1mm]
  \caption{The variation of granulation RMS contrast with $r_0$ for
    MFBD restored images. Plus-symbols show the contrast after
    conventional MFBD processing, squares the result after $\hat
    S$-compensation. Red and blue symbols correspond to 630~nm and
    538~nm resp., with the contrasts at 630~nm multiplied by a factor
    1.22. The upper panel shows the results from processing of images
    in blocks of 2.2~sec., the bottom panel from MFBD processing of the entire scan to determine the wavefronts, but with restoration of the images and high-order compensation carried out in 2.2~sec. blocks. The dotted curve is identical to that shown in the bottom panel of Fig.\@~\ref{fig:scatter_r0_raw}.}
  \label{fig:contrast.100}
\end{figure}

Granulation contrast was measured over the 200$\times$400-pixel
subfield shown in Fig.\@~\ref{fig:restored}. This region appears
reasonably free from strong magnetic fields, as judged by the near
absence of bright points and other sub-granular structure.

The measured contrasts (blue: at 538~nm; red: at 630~nm, multiplied by
1.22) are shown in Fig.\@~\ref{fig:contrast.100}. The upper panel
shows the contrasts measured from blocks of images recorded during
2.2~sec intervals. The contrasts plotted in the bottom panel
correspond to MFBD sets based on full scan data. The plus symbols
refer to contrasts obtained after conventional MFBD processing with
100 KL aberrations, the squares to contrasts obtained after
compensation also of high-order aberrations. We note that the results
are quite similar for the two types of MFBD processing when $r_0$ is
large, as expected. Data processed with 2.2~sec sets show a slow and
gradual decrease of the RMS contrast with decreasing $r_0$ after
high-order compensation (squares). Inspection of the restored images
shows that this decrease in contrast is associated with a small but
noticeable decrease of image quality. For values of $r_0$ smaller than
9--10~cm, this degradation is obvious as reduced spatial resolution. A possible explanation for the reduced RMS contrast in poor seeing is the finite integration time used: A seeing layer moving at 10~m\,s$^{-1}$ is displaced by 16~cm peak-to-valley (5~cm RMS) during a 16~ms exposure and will cause increased smearing of the wavefront with decreasing $r_0$.

For the full scan data sets obtained in strongly varying seeing
conditions, the trend is quite the opposite to that seen with the 2.2~sec data sets. For these data, the RMS
contrast \emph{increases} with decreasing $r_0$ after $\hat S$
compensation. This is again explained by the compensation effects
discussed in Sect.\@~\ref{sec:without-shat}: MFBD processing assumes
that a unique object $F$ (see Eq.\@~\ref{eq:2}) is ``responsible''
for all observed images of a data set, and that this object can be
estimated with a fixed number of aberration parameters for each of the
observed images. In stable seeing, this works well. However, when the
seeing is strongly variable, this causes inconsistencies. During
moments of bad seeing, the missing high-order aberrations lead to such
poorly represented wavefronts that the observed images are
inconsistent with the images recorded in good seeing. MFBD compensates
this inconsistency by increasing the amplitudes of the
\emph{low-order} aberrations in bad seeing. When compensating for the
high-order aberrations in the final reconstruction of the images, the
overestimated wavefront RMS leads to contrast values that are too high
when $r_0$ is small. Comparison between the restored images from the
2.2~sec and full scan MFBD shows no differences in image quality apart
from the differences in contrast; quite clearly either way of MFBD
processing works very well as long as the number of images in the data
set is not too small.

\subsubsection{Comparison with 3D MHD simulations}

Based on the results of the 2.2~sec data sets, we are finally in a
position to compare the measured granulation contrasts with those of
simulations. The highest measured RMS contrast is 9.8\% at 538~nm and
8.5\% at 630 nm (note that the contrast values at 630~nm in
Fig.\@~\ref{fig:contrast.100} are multiplied by 1.22).

To compare measured RMS contrasts with those of the 3D simulations, we
calculated a synthetic spectrum covering the \ion{Fe}{i} 630.1~nm and
630.2~nm lines and nearby continuum, using the same simulation
snapshot shown in Fig.\@~\ref{fig:images} but for a heliocentric
distance corresponding to $\mu=0.87$. Similarly, we included the weak
\ion{C}{i} line and two stronger \ion{Fe}{i} lines for calculating a
synthetic spectrum corresponding to the 538~nm observations. We
multiplied the synthetic spectra obtained at each pixel with the
transmission profiles of the CRISP pre-filters used to record the
observed data and calculated the contrast of the granulation pattern.
We obtained an RMS contrast of 13.4\%, compared to 13.9\% at a clean
nearby continuum wavelength, for the 630~nm data and 17.3\%, compared
to a continuum contrast of 17.8\% for the 538~nm data. In comparison,
the corresponding values at $\mu=1$ are 14.2\% and 14.6\% at 630~nm
and 18.2\% and 18.6\% at 538 nm. It should be emphasized that these
RMS contrasts have been calculated without degrading the spatial
resolution of the synthetic data to that of a 1-m telescope, nor have
any effects of noise on the data been included.

The measured RMS contrasts are only 57\% at 538~nm and 63\% at 630~nm
of those expected, clearly demonstrating the existence of sofar 
unidentified stray light sources and suggesting that this stray light
increases at shorter wavelengths. The origin of this stray light will
be investigated in forthcoming papers.

\subsubsection{Wavefronts}
\label{sec:wavefronts}

\begin{figure}[!t]
  \centering
  \includegraphics[bb=15 11 493 343,clip,width=\linewidth]{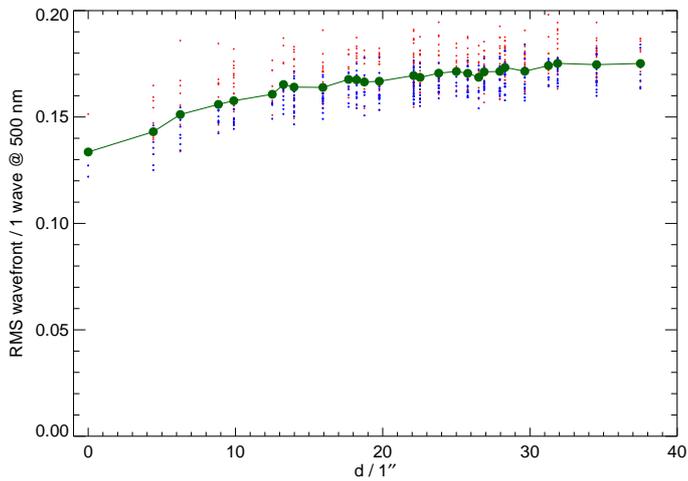}
  \caption{The variation of the wavefront RMS (in units of waves at
    500~nm) with distance $d$ from FOV center in arc seconds. MFBD
    restorations with $M=100$ corresponding to the bottom tile of
    Fig.\@~\ref{fig:contrast.100}. Blue: two sets of 538~nm data; Red:
    two sets of 630~nm data; Green: average of all four sets.}
  \label{fig:wavefrontrms}
\end{figure}

Figure~\ref{fig:wavefrontrms} shows the variation of the wavefront RMS
with distance from the center of the FOV, estimated from MFBD
processing. The plot corresponds to two good data sets at each
wavelength. Wavefront tip-tilt (KL coefficients 2 and 3) are not
included in the calculation. As expected, the wavefront RMS is
smallest at the center of the FOV ($d=0$), where the AO is locked, and
gradually increases away from the lock point. This demonstrates that
in contrast to speckle techniques, MOMFBD techniques can be used with
AO compensated images to retrieve wavefronts without any a priori
knowledge of how anisoplanatism is introduced by the AO system and the
Earth's atmosphere.

\begin{figure}[!t]
  \centering
  \includegraphics[bb=10 9 493 343,clip,width=\linewidth]{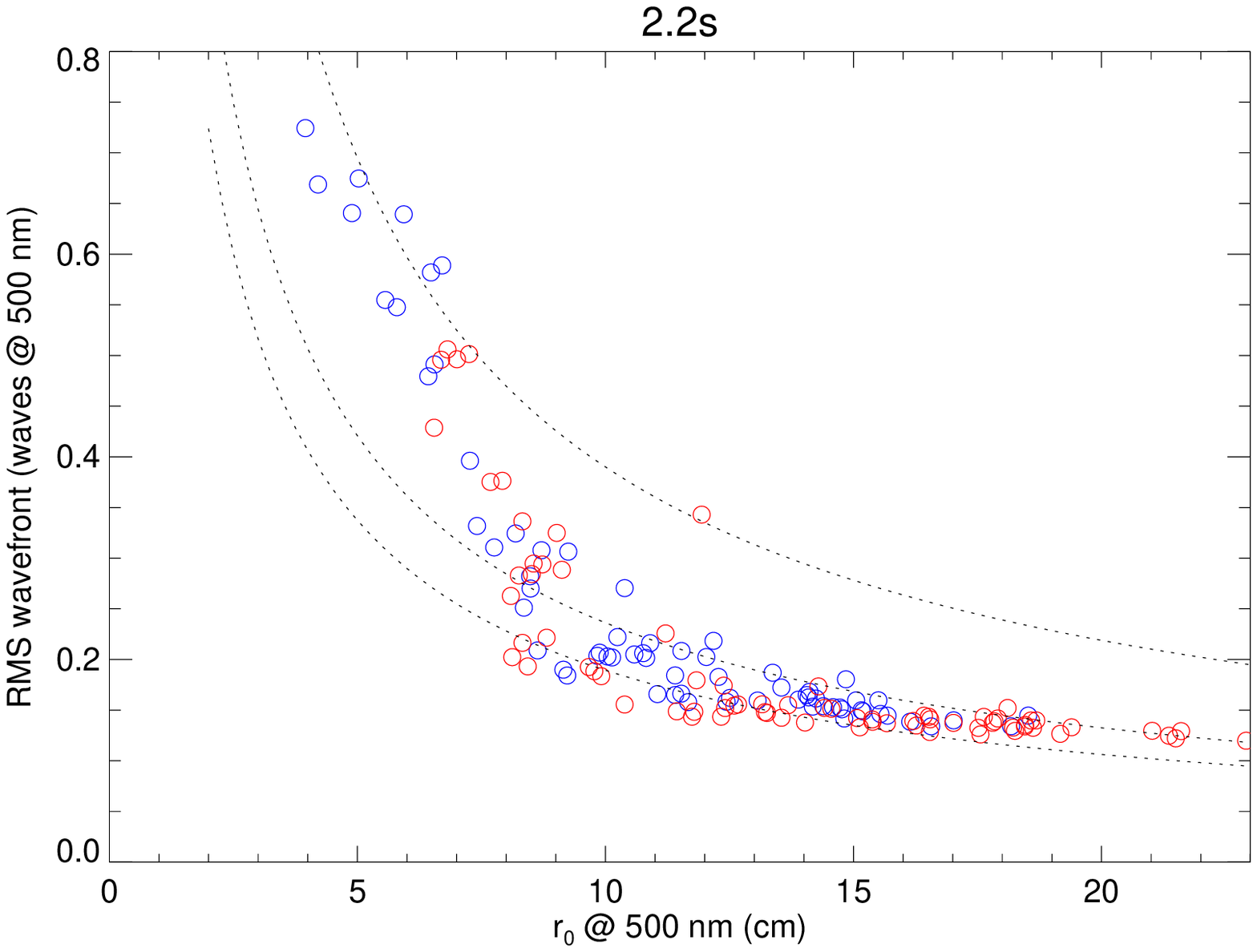}\\[1mm]
  \includegraphics[bb=10 9 493 343,clip,width=\linewidth]{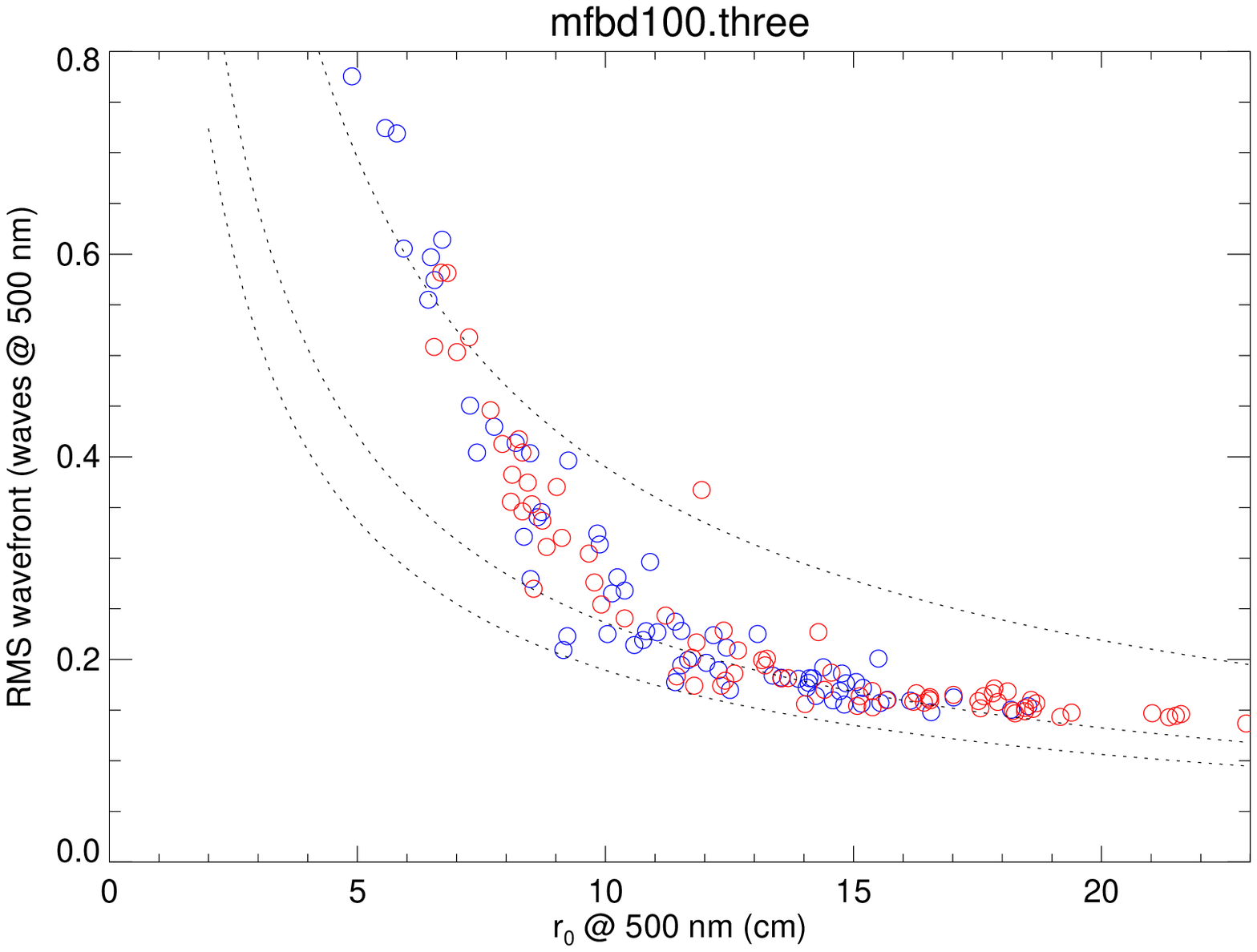}\\[1mm]
  \caption{RMS wavefronts vs $r_0$. Red and blue circles: RMS
    wavefronts as estimated with MFBD using $M=100$ modes and 2.2~sec
    data blocks (top tile), and 12--17~sec data blocks (bottom tile).
    Black dotted curves: Wavefront RMS expected from perfect
    correction of the first $N-1$ modes for $N=3$, 10, 17 (top to
    bottom) KL modes.}
  \label{fig:r0_WFRMS}
\end{figure}

Figure \@~\ref{fig:r0_WFRMS} shows the variation of the MFBD estimates
of the wavefront RMS with $r_0$ at the center of the FOV,
corresponding to the AO lock-point. The upper panel shows the
wavefront RMS obtained by MFBD processing of data in 2.2~sec blocks,
the bottom panel by processing full scan data. The three dotted curves
correspond to the expected wavefront RMS with the AO system perfectly
compensating the first 2, 9 and 16 KL aberrations (top to bottom,
corresponding to $N=3$, 10 and 17 in Eq.\@~(\ref{eq:9})). The top
panel shows that the MFBD estimate of the wavefront RMS is consistent
with perfect AO compensation of about 15 KL modes when $r_0$ is
approximately 10--12~cm, whereas when $r_0$ is larger than 20~cm, the RMS
wavefront approaches the values expected with perfect compensation of
only 10 KL modes. We conclude that the MFBD wavefront RMS estimates
are similar, although somewhat higher, than what we expect from the
independent WFWFS estimates of $r_0$. This strongly suggests that the
WFWFS measurements of $r_0$ are at most associated with small
systematic errors.

For values of $r_0$ smaller than about 9.5~cm, the MFBD estimated
wavefront RMS shows a rapid transition to values equivalent to only 2
perfectly corrected KL modes. This corresponds to pure tip-tilt
correction, or recording images with short exposures as used with the
present data, but with the AO system not functioning at all. Such
failure of solar AO systems to lock in poor seeing is well-known: When
$r_0$ is smaller than the sub-aperture diameters of the Shack-Hartmann
wavefront sensor controlling the AO system, the granulation images
start to degrade, leading to image positions that are poorly
determined with cross-correlation techniques. The SST AO system uses
fairly large sub-aperture diameters, 14~cm, and is more vulnerable to
bad seeing than other solar AO systems. The data in
Fig.\@~\ref{fig:r0_WFRMS} suggest that the efficiency of the SST AO
system starts to degrade when the ratio of $r_0$ to the sub-aperture
diameter is about 0.7. This ratio is similar to what was found for the
WFWFS, using sub-aperture diameters of about 9.8~cm, leading to
strongly increased noise levels when $r_0$ is less than about 7~cm
\citep[][]{2010A&A...513A..25S}.
 
The bottom panel in Fig.\@~\ref{fig:r0_WFRMS}, based on full-scan data
recorded in strongly variable seeing, shows a variation of wavefront
RMS with $r_0$ that is quite similar to that of the upper panel.
However, the wavefront RMS, in particular during bad seeing, is
systematically higher for the full-scan data. 
This is consistent with the systematic differences in granulation contrasts found for these data, discussed in the previous section: In strongly varying seeing the limited number of aberrations used to model the wavefront becomes a problem in particular during moments of poor seeing. The MFBD compensates this by increasing the RMS of the low-order aberrations. When combined with $\hat S$-compensation, this leads to overcompensation. At the same time, MFBD optimization is dominated by the \emph{best} frames, because of the weighting of the transfer functions used to restore the object, see Eq.\@~\ref{eq:4}. This reduces the contribution of the poorer and overcompensated frames in the restored image and thus leads to a reduction of the overcompensation effects when images recorded in \emph{variable} seeing are combined and restored.

\section{Stray-Light compensation}
\label{sec:strayl-corr}

An in-depth discussion of stray light measurements and compensation is beyond the scope of the present paper. However, we note that the primary focal
plane used to calibrate the control matrix of an AO system can also be
used to aid stray light calibrations of any following optics and
instrumentation. Such (partial) stray light calibrations can be made
by locking the AO system on the stray light target and recording
images with the science camera at several focus positions. Processing
these images with JPDS techniques, allows the effects of any residual
aberrations to be separated from those of stray light, making such
measurements robust.

Spatial stray light can be simply modeled as a convolution:
\begin{equation}
  D_k = F T_s T_k + N_k ,
  \label{eq:13}
\end{equation}
where $T_s$ is the Fourier transform of the stray light PSF, $t_s$.
The residual aberrations can be identified and determined
by recording pinhole images at different focus locations and the corresponding
transfer function $T_k$ can be calculated and compensated for with phase diversity techniques. The stray light transfer function $T_s$ is assumed to be invariant with respect to the focus position. It is
then evident that phase diversity image restoration will produce a restored ``object'' that is the product of $F$ and $T_s$ and that $T_s$ can therefore only be determined if the object $F$ is known. For example, using
a circular pinhole as object, corresponds to a binary object that
is of constant intensity within a radius $r$ and zero outside that radius. Recording images of pinholes of different diameters should allow the
stray light PSF to be determined with confidence. Procedures
for such stray light measurements and compensation will be described
in a forthcoming paper.\footnote{We remark that such a calibration
  will also include the detector MTF. This MTF is primarily the
  combination of two effects: The first is from the integration over
  the finite size of the pixels and the second is from charge
  diffusion in the silicon substrate of the CCD detector \cite[e.g.,
  ][]{stevens94analytical}. With back-illuminated thinned CCDs
  operating at near infrared wavelengths, as used with CRISP, there is
  a third stray light source from light that is first transmitted
  through the thin silicon layer of the CCD and then scattered back
  into the beam. Compensation for such stray light needs special
  calibration and compensation techniques (de la Cruz Rodr\'iguez et al., in preparation).}

Of relevance to the present paper is that the stray light PSF, as well
as $\hat S$ in bad seeing, have wings that can extend well beyond
several arc sec. Proper compensation for such a broad PSF during
MFBD/JPDS processing would cause problems with the small sub-fields
needed to deal with anisoplanatism. The calculation of $\hat S$
involves atmospheric transfer functions modified by high-order
aberrations but does not involve the observed images $d_k$. A
reasonable expectation is that $\hat S$ should show relatively small
and smooth variations over a large FOV. This would allow $\hat S T_s$,
averaged over a large fraction over the FOV, to compensate for the
combined effects of high-order atmospheric aberrations and stray light
using only a few deconvolutions, independent of the sub-fielding used
in previous MFBD or JPDS processing. The restored images can then be
properly compensated for stray light originating from the broad wings
of the PSF corresponding to $\hat S T_s$.

\section{Conclusions}
\label{sec:conclusions}

The development of techniques for restoration of solar images based on multi-frame blind deconcolution (MFBD) such as phase diversity \cite[PD;][]{lofdahl94wavefront}, joint phase diversity speckle \cite[JPDS;][]{paxman92joint} and multi-object multi-frame blind deconvolution \cite[MOMFBD;][]{lofdahl02multi-frame,noort05solar}
is of major importance to present and future broad-band imaging, imaging spectrometry and imaging spectropolarimetry with ground-based solar telescopes. An important advantage of these techniques is that they do not rely on statistical properties of atmospheric seeing, including anisoplanatism introduced by the Earth's atmosphere and a conventional or multi-conjugate AO system. These techniques, implemented in the MOMFBD C++ code developed by \citet{noort05solar}, have been used extensively to process data sets recorded with the SST. This has
resulted in stunning time sequences with near-diffraction limited
resolution. The remarkably stable image quality achieved when many
exposed frames are processed as a single data set strongly indicates
that this code is working well. The present work represents the first
systematic evaluation of its performance.

Our simulations
confirm and explain the excellent performance of the code with images
recorded using a low-order (37 electrode) AO system. In particular, we
demonstrate that even though MFBD processing of AO compensated images
leads to poorly determined wavefronts, this is of secondary importance.
The transfer functions, needed to restore images, are actually better
determined than expected from the number of KL modes used to represent
the wavefront. A surprising result is that the addition of a phase
diversity (defocused) channel not necessarily improves the result when many images are processed simultaneously: adding such a channel constrains the \emph{wavefronts} to more closely represent reality, but the obtained \emph{transfer functions} may actually be more accurate without the defocused channel. From this perspective, MFBD processing rather than JPDS processing is preferrable. However, when applying $\hat S$-compensation to the restored images, the more accurate wavefronts of JPDS constitute an advantage. A disadvantage of JPDS processing with an added diversity channel is that defocused images are fuzzier, causing ``leakage'' of (stray-)light across
subfield boundaries and wrapping around when using FFTs to perform
convolutions. This degrades the quality of the restorations or even cause the inversions to fail when the subfields are small. We can thus not simply conclude that the MFBD or JPDS method of processing in general is preferrable, but always \emph{recording} also defocused images is a good strategy since this allows either mode of processing the data.

A drawback of MFBD and JPDS techniques is that
truncation of the wavefront expansion is necessary. This leaves a tail
of uncompensated high-order aberrations and under-compensation of the
restored images. Based on restorations of images recorded in strongly
variable seeing, we have found that MFBD processing compensates for
the missing high-order aberrations by increasing the amplitudes of the
low-order aberrations. This compensation effect is particularly large
during moments of bad seeing but the corresponding exposed frames are given relatively small weight in the restored images. This explains the apparent stable image quality of movies made from MFBD processed data.

We have proposed a method for further compensation of missing
high-order aberrations, based on statistics of Kolmogorov turbulence.
Our simulations show that such compensation in principle should allow
\emph{perfect} image restorations in good to excellent seeing conditions. A
major advantage of the method is that very good results can be
obtained even when only a small number of frames are used to restore an object.
However, the high-order mode compensations already
introduced by MFBD or JPDS image processing limits the overall
improvement factor to about 2--3, which nevertheless should represent
an important improvement of the photometric quality of the restored
images. A problem is MFBD processing of data sets
recorded during strongly variable seeing. In such conditions, the use of a truncated wavefront leads to overcompensation for the images recorded during
moments of poor seeing such that the contrast of the restored images
increases with decreasing $r_0$ when applying $\hat S$-compensation. By breaking up the processing in smaller data sets, these overcompensation effects are strongly reduced.

Using images recorded with the SST in 1\arcsec{} seeing, we have
calculated the RMS contrast of solar granulation. The obtained ``raw'' contrast values range from 7.5\% at 538~nm and 6.5\% at 630~nm for the best
images, observed away from sun center (at $\cos\theta = \mu = 0.87$)
and through pre-filters that contain a few moderately strong \ion{Fe}{i}
lines. Due to the previous compensation by the AO system, the measured
RMS contrast increases by only a fairly small amount to 9\% and 7.5\%
respectively for MFBD processed images using 100 KL aberrations.
Simultaneous measurements of Fried's parameter $r_0$ were used to aid
compensation for high-order aberrations and yields an increase of the
RMS contrast to close to 10\% at 538~nm and 8.5\% at 630~nm. The
corresponding RMS contrast of synthetic images obtained from 3D MHD
simulations, taking into account the strongest lines within the
filters used and an inclination angle of $\theta=30\degr$, is
17.4\% at 538~nm and 13.3\% at 630~nm. This implies a remaining
discrepancy of 37--43\% of the true contrast, the major part of which must be from stray light in the Earth's atmosphere, the telescope, the subsequent
re-imaging optics, the finite (16~ms) integration time used to record
the images and/or anisoplanatism, scintillation and image scale variations from high-altitude seeing. The origin of this missing stray light will be the topic of forth-coming papers.

\begin{acknowledgements}
  Part of this work was supported by a Marie Curie Early Stage Research Training Fellowship of the European Community's Sixth Framework Programme under contract number MEST-CT-2005-020395: The USO-SP International School for Solar Physics. This work was also partially supported by the European Commission
through the collaborative project  212482 'EST: the large aperture European
Solar Telescope´ Design Study (FP7 - Research Infrastructures) and by a planning grant from the Swedish Research Council.

M.G.L. is grateful to George
  (Guang-ming) Dai for sharing the his calculated radial Karhunen--L\`oeve functions, used for the atmospheric turbulence
  simulations. The Swedish 1-m Solar Telescope is operated on the
  island of La Palma by the Institute for Solar Physics of the Royal
  Swedish Academy of Sciences in the Spanish Observatorio del Roque de
  los Muchachos of the Instituto de Astrof\'isica de Canarias.
\end{acknowledgements}


\end{document}